\documentclass[final,12pt]{clear2024} 

\usepackage[ruled]{algorithm2e} 
\usepackage{footnote}
\usepackage{algorithm, algpseudocode}
\usepackage{times}
\newcommand{\ci}{\perp\!\!\!\perp}
\newcommand{\indep}{\perp \!\!\! \perp}
\newcommand{\tailhead}{{\rightarrow}}
\newcommand{\headtail}{{\leftarrow}}
\newcommand{\headhead}{{\leftrightarrow}}
\newcommand{\ohead}{{\circ\!{\rightarrow}}}
\newcommand{\heado}{{{\leftarrow}\!\circ}}
\newcommand{\oo}{{\circ\!{\--}\!\circ}}
\newcommand{\xx}{{\times\!{\--}\!\times}}

\title[Bootstrap aggregation for time series causal discovery]{Bootstrap aggregation and confidence measures to improve time series causal discovery}

\clearauthor{%
 \Name{Kevin Debeire} \Email{kevin.debeire@dlr.de}\\
 \addr Deutsches Zentrum f\"{u}r Luft- und Raumfahrt (DLR), Institut f\"{u}r Physik der Atmosph\"{a}re, Oberpfaffenhofen, Germany.\\
 Deutsches Zentrum f\"{u}r Luft- und Raumfahrt (DLR), Institut für Datenwissenschaften, Jena, Germany.
 \AND
\Name{Andreas Gerhardus}$^*$ \Email{andreas.gerhardus@dlr.de}\\
 \addr Deutsches Zentrum f\"{u}r Luft- und Raumfahrt (DLR), Institut für Datenwissenschaften, Jena, Germany.
 \AND
 \Name{Jakob Runge}$^*$ \Email{jakob.runge@dlr.de}\\
 \addr Deutsches Zentrum f\"{u}r Luft- und Raumfahrt (DLR), Institut für Datenwissenschaften, Jena, Germany. \\
 Technische Universität Berlin, Faculty of Computer Science, Berlin, Germany.
 \AND
 \Name{Veronika Eyring} \Email{veronika.eyring@dlr.de}\\
 \addr Deutsches Zentrum f\"{u}r Luft- und Raumfahrt (DLR), Institut f\"{u}r Physik der Atmosph\"{a}re, Oberpfaffenhofen, Germany.\\
 University of Bremen, Institute of Environmental Physics (IUP), Bremen, Germany.\\
 \\
$^*$These corresponding authors contributed equally.
}

\begin{document}

\maketitle

\begin{abstract}%
Learning causal graphs from multivariate time series is an ubiquitous challenge in all application domains dealing with time-dependent systems, such as in Earth sciences, biology, or engineering, to name a few. Recent developments for this causal discovery learning task have shown considerable skill, notably the specific time-series adaptations of the popular conditional independence-based learning framework. However, uncertainty estimation is challenging for conditional independence-based methods. Here, we introduce a novel bootstrap approach designed for time series causal discovery that preserves the temporal dependencies and lag-structure. It can be combined with a range of time series causal discovery methods and provides a measure of confidence for the links of the time series graphs. Furthermore, next to confidence estimation, an aggregation, also called bagging, of the bootstrapped graphs by majority voting results in bagged causal discovery methods. In this work, we combine this approach with the state-of-the-art conditional-independence-based algorithm PCMCI+. With extensive numerical experiments we empirically demonstrate that, in addition to providing confidence measures for links, Bagged-PCMCI+ improves in precision and recall as compared to its base algorithm PCMCI+, at the cost of higher computational demands. These statistical performance improvements are especially pronounced in the more challenging settings (short time sample size, large number of variables, high autocorrelation). Our bootstrap approach can also be combined with other time series causal discovery algorithms and can be of considerable use in many real-world applications.
\end{abstract}


\section{Introduction}

\label{intro}
Since a rigorous mathematical framework for causal inference has been established in the seminal works of Pearl, Spirtes, Glymour, Scheines, and Rubin \citep{imbens_rubin_2015,pearl_2009,Spirtes2000-SPICPA-2,Rubin1974EstimatingCE}, causal inference has undergone continuous  developments to address the challenges of real-world problem settings.
Learning causal graphs from data (termed causal discovery) is a main pillar of causal inference and of high interest in many fields where not even qualitative causal knowledge in the form of graphs is available, such as in biology \citep{Friedman2000UsingBN}, neuroscience \citep{Kaminski2001EvaluatingCR}, or Earth system sciences \citep{ebert2012causal,Runge2019InferringCF,kretschmer2016using,Galytska_2022,Karmouche2023RegimeorientedCM}. Furthermore, data in these fields typically comes in the form of time series, constituting a more general problem setting than the standard \emph{i.i.d.}-case. In \citep{Runge2019InferringCF}, the authors give an overview of a few main categories of \textbf{time series causal discovery} methods: Granger causality and its extensions \citep{granger_2001}, nonlinear state-space methods (CCM \citep{Sugihara2012DetectingCI}), causal network learning algorithms (for example, naive adaptations of the PC-algorithm \citep{Spirtes1991AnAF}, PCMCI and its extensions PCMCI+ \citep{runge19ScienceAdv,Runge2020DiscoveringCA,NEURIPS2020_94e70705}, FCI \citep{Spirtes2000-SPICPA-2,Zhang2008OnTC}), Bayesian score-based approaches \citep{Chickering1996LearningEC}, and the structural causal model framework (e.g., VarLiNGAM \citep{JMLR:v11:hyvarinen10a}).
While the state-of-the-art time series causal discovery methods have seen considerable improvements over the years, this is still a very challenging task. A particular drawback in practical applications is that most methods output single graphs without the option to assess the uncertainty or confidence in the causal links. Approaches in this direction are based on taking the absolute p-value over all conditioning sets \citep{Strobl2016EstimatingAC}, which presents a conservative option and does not permit to assess the uncertainty in orientations.

Independent of the development of causal discovery methods, \citet{Breiman1996BaggingP} introduced \textbf{bootstrap aggregation} (bagging) which has initially been used to improve the accuracy and stability of machine learning algorithms. In bagging, a random sample in the training set is selected with replacement —meaning that each data point can be drawn more than once. Several data samples are generated in this fashion to produce a set of replicates (also called resamples). The machine learning models are then trained independently on each replicate and, finally, the outputs are averaged for prediction tasks or aggregated for classification tasks (for example by majority voting). Since its introduction, bagging has been extensively used in combination with other machine learning algorithms \citep{Dietterich2000,GALAR2012,GANAIE2022105151}.

Combining bagging and causal graphical model algorithms has been proposed to improve the stability of graphical model learning \citep{Meinshausen2008StabilityS,Li2011BOOTSTRAPIF}, as the estimation of graphical models is relatively sensitive to small changes of the original data. For example, \citet{Wang2014LearningDA} introduce an aggregation approach of directed acyclic graphs (DAGs) by minimizing the overall distance of the aggregated graph (based on structural hamming distance) to the ensemble of DAGs. Bootstrapping a causal discovery algorithm and returning a summary graph constructed by a voting scheme is a feature of the TETRAD project \citep{Ramsey2018TETRADA}. \citet{GUO2021} propose a two-phase causality ensemble framework to combine results from different data partitions (instead of bootstrap samples) and causal discovery algorithms into a single output graph using majority voting. In addition, the idea of measuring the uncertainty or confidence for an edge of an estimated graph from the edge frequency based on the graphs learned on bootstrap samples has been suggested in \citet{Friedman1999DataAW,Imoto2002BootstrapAO,Mooij2016JointCI}. However, none of these approaches is directly transferable to time series causal discovery because their bootstrap sampling needs to be adapted to the lagged interdependencies of time series data.

Our \textbf{main contribution} is the introduction of a bootstrap method for time series causal discovery which preserves temporal dependencies. Our method allows (1.) to obtain \textbf{uncertainty estimates} for the links of the output graph, and (2.) \textbf{improves the stability and accuracy} by aggregating the ensemble of bootstrap graphs to one single output graph with majority voting at the level of each individual edge. In principle, our method can be paired with any time series causal discovery algorithm. In the main text, we investigate the combination with PCMCI+~\citep{Runge2020DiscoveringCA} as a representative of a state-of-the-art constraint-based time series causal discovery method. Results for further methods are presented in the Appendices. \textbf{The paper is structured as follows}. In Section~\ref{section:causal_discovery}, we give an overview of time series causal discovery and the PCMCI+ method. In Section~\ref{section:bagging}, we present our bagging and confidence measure technique which we combine with PCMCI+ (Bagged-PCMCI+). With a range of numerical experiments, we show in Section~\ref{section_num_exp} that Bagged-PCMCI+ outperforms standard PCMCI+ and that our method to measure confidence for links is effective. Finally, we summarize the paper in Section~\ref{section:ccl}. The paper is accompanied by an Appendix.

\section{Time series causal discovery}
\label{section:causal_discovery}
\subsection{Preliminaries}

We consider discrete-time structural causal processes \(\mathbf{X}_t = (X_t^1, ..., X_t^N)\) such that 
\begin{equation}
\label{eq1}
X_t^j := f_j({\rm pa}(X_t^j),\eta_t^j) \quad \forall j \in \{1, \dots, N\} \quad \forall t.
\end{equation}
Here, \(f_j\) are arbitrary measurable functions that depend non-trivially on all their arguments 
and \(\eta_t^j\) are mutually and temporally independent noises. In a time series graph \(\mathcal{G}\), the nodes represent the variables \(X_t^j\) at different time lags. The \emph{causal parents} \({\rm pa}(X_t^j)\) are the set of variables on which \(X^j_t\) depends, and a causal link from \(X^i_{t-\tau}\) to \(X^j_t\) exists if \(X^i_{t-\tau} \in {\rm pa}(X_t^j) \) for a time lag \(\tau\). 
A link \(X^i_{t-\tau} \rightarrow X^j_t\) is a called \emph{lagged} if \(\tau > 0\), else it is called \emph{contemporaneous}. 
In this work, we assume \emph{stationarity} of the causal links: that is, if the causal link \(X^i_{t-\tau} \rightarrow X^j_t\) exists for some time \(t\), then \(X^i_{t'-\tau} \rightarrow X^j_{t'}\) also exists for all times \(t' \neq t\). We define the set \(\mathcal{A}(X^j_t)\) of non-future adjacencies of variable \(X_t^j\) as the set of all variables \(X_{t-\tau}^i\) for \(\tau \geq 0 \) that have a causal link with \(X_t^j\).

\subsection{PCMCI+}
We focus on the combination of our bagging approach with PCMCI+ as it is a widely-used state-of-the-art algorithm for the setting it considers, that is, lag-resolved time series causal discovery with contemporaneous edges but without hidden confounders.
PCMCI+ learns the causal time series graph including lagged and contemporaneous links (up to Markov equivalence) under the standard assumptions of Causal Sufficiency, Faithfulness, and the Causal Markov condition, as well as causal stationarity \citep{Runge2020DiscoveringCA}. To increase the detection power and maintain well-calibrated tests, PCMCI+ optimizes the choice of conditioning sets in the conditional independence (CI) tests. It is based on two central ideas: (1)~separating the skeleton edge removal phase into a lagged and contemporaneous conditioning phase, and (2) constructing conditioning sets in the contemporaneous conditioning phase via the so-called momentary conditional independence (MCI) approach~\citep{runge2019detecting}, as explained below. Moreover, PCMCI+ is order-independent~\citep{Colombo2014OrderIndependence}, which implies that the output does not depend on the order of the variables \(X^j\). More details and examples of PCMCI+ can be found in \cite{Runge2020DiscoveringCA}, here we briefly summarize it.

In its \textbf{first phase}, PCMCI+ starts with a fully connected graph and then removes adjacencies among variable pairs by conditional independence testing. 
To this end, the PC\(_1\) algorithm tests all lagged pairs $(X^i_{t-\tau}, X_t^j)$ for $\tau > 0$ conditioning on subsets $\mathbf{S}_k\subseteq \mathcal{A}(X^j_t) \cap \mathbf{X}^-_t$ with the lagged variables $\mathbf{X}^-_t=(\mathbf{X}_{t-1}, \mathbf{X}_{t-2},\ldots, \mathbf{X}_{t-\tau_{\max}})$ up to a maximum time lag $\tau_{\max}$. If (conditional) independence is detected, the adjacency is removed. The subsets $\mathbf{S}_k$ are chosen with increasing cardinality $k$: For $k=0$ all \(X^i_{t-\tau}\) with \(X^i_{t-\tau} \ci X^j_t\) are removed, for \(k=1\) those with \(X^i_{t-\tau} \ci X^j_t | \mathbf{S}_1{\setminus} X^i_{t-\tau}\) where \(\mathbf{S}_1\) is the adjacency with largest association (not counting $X^i_{t-\tau}$) with $X^j_t$ from the previous step, for \(k=2\) those with \(X^i_{t-\tau} \ci X^j_t | \mathbf{S}_2{\setminus} X^i_{t-\tau}\) where \(\mathbf{S}_2\) are the two adjacencies with largest association (not counting $X^i_{t-\tau}$) with $X^j_t$ from the previous step, and so on. Association strength is measured by the absolute test statistic value of the CI test. This procedure improves recall and speeds up the skeleton phase as compared to the standard PC algorithm skeleton phase. The resulting lagged adjacency sets for each $X^j_t$ of this first phase are denoted \(\mathcal{B}^-_t(X_t^j)\). In its \textbf{second phase}, the graph \(\mathcal{G}\) is initialized with all
contemporaneous adjacencies plus all lagged adjacencies
from \(\mathcal{\hat{B}}^-_t(X_t^j)\) for all \(X_t^j\) found in the PC\(_1\) algorithm in the first phase. The second phase of PCMCI+ tests all, contemporaneous and lagged, adjacent pairs \((X^i_{t-\tau}, X_t^j)\) for \(\tau\geq 0\), but iterates only through contemporaneous conditions \(\mathbf{S} \subseteq  \mathcal{A}(X^j_t) \cap \mathbf{X}_t\) with the MCI test
\(X^i_{t-\tau} \indep X_t^j \mid \mathbf{S},\mathcal{\hat{B}}^-_t(X_t^j) \setminus \{X^i_{t-\tau} \}, \mathcal{\hat{B}}^-_t(X^i_{t-\tau})\).
The conditioning on \( \mathcal{\hat{B}}^-_t(X_t^j)\) blocks paths through lagged parents while the conditioning on \(\mathcal{\hat{B}}^-_t(X^i_{t-\tau}) \) has been shown to lead to well-calibrated tests even for highly autocorrelated time series~\citep{runge2019detecting,Runge2020DiscoveringCA}. After these tests, the time-series-adapted collider orientation phase and rule orientation phase are applied: the former rule orients the collider motifs that contain contemporaneous links based on unshielded triples while the latter rule orients the remaining contemporaneous links based on the Meek rules \citep{meek1995}.

In the final graph, the following link types can connect a pair $(X^i_{t-\tau}, X^j_t)$ of vertices: no link (i.e., pair is non-adjacent), direct link $X^i_{t-\tau} \tailhead X^j_t$, opposite direct link $X^i_{t} \headtail X^j_t$ (only for $\tau = 0)$, unoriented link $X^i_t \oo X^j_t$ (only for $\tau = 0$), conflict-indicating link $X^i_t \xx\ X^j_t$ (due to finite sample effects or violations of assumptions, only for $\tau = 0$). 

Like other CI-based methods, PCMCI+ has the free parameters \(\alpha_{PC}\) (significance level of CI tests), \(\tau_{\max}\) (maximal considered time lag), and the choice of the CI test. \(\alpha_{PC}\) in PCMCI+ turned out empirically to be an upper bound on the false positives. As opposed to such a statistically-motivated choice, it can also be chosen based on cross-validation or an information criterion \citep{Runge2020DiscoveringCA}. \(\tau_{\max}\) should be larger or equal to the maximum assumed true time lag of any parent and can in practice also be chosen based on model selection. However, the numerical experiments indicate that a too large \(\tau_{\max}\) does not degrade performance much \citep{Runge2020DiscoveringCA}.
PCMCI+ can flexibly be combined with different CI tests for nonlinear causal discovery, and for different variable types (discrete or continuous, univariate or multivariate).

\section{Bootstrapping time series causal discovery}\label{section:bagging}
In this section, we motivate and explain our technical contributions to bagging and bootstrap uncertainty quantification of time series causal discovery methods.


\subsection{Motivational example}

\begin{figure}[ht]
  \centering
  \includegraphics[width=0.77\linewidth]{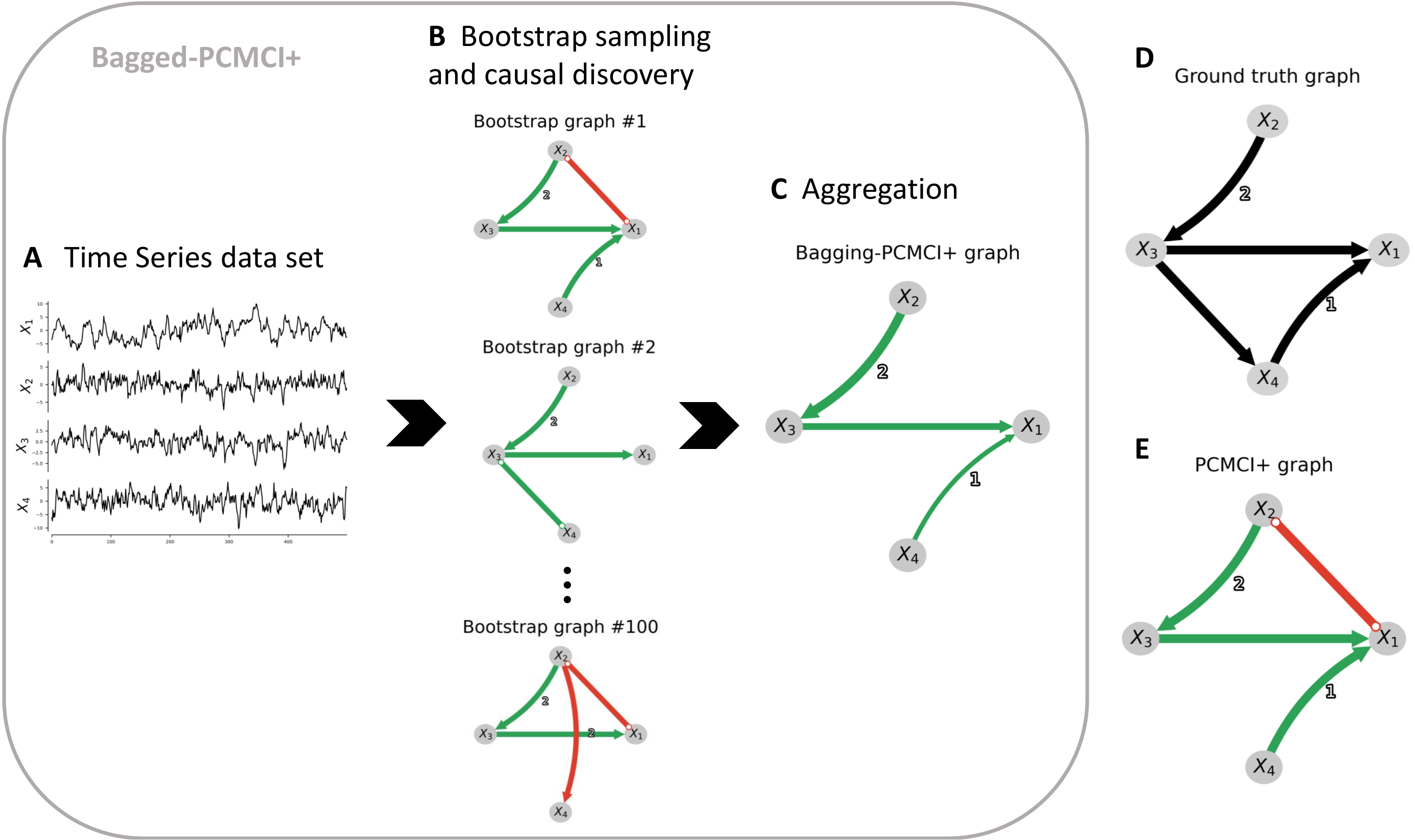}
  \caption{Motivational example and schematic of our approach. \textbf{(A)} Time series for a model as in eq.~\eqref{eq1} with linear functions $f_j$. \textbf{(B)} Ensemble of $B=100$ estimated causal graphs obtained by applying PCMCI+ to each bootstrap dataset randomly drawn as described in Sec.~\ref{sec:sampling-approach}. Green links indicate true positives, red links represent false positives, and the numbers next to the edges indicate non-zero time lags. \textbf{(C)} Graph obtained by aggregating the ensemble of graphs as explained in Sec.~\ref{sec:aggregation-approach}, with confidence scores obtained as explained in Sec.~\ref{sec:confidence-approach} and visualized as the thickness of edges. \textbf{(D)} Ground truth causal graph of the model. \textbf{(E)} Estimated causal graph obtained by applying PCMCI+ to the data in part \textbf{A}.}
  \label{flowchart}
\end{figure}

\textbf{Figure \ref{flowchart}} illustrates the benefits and overall structure of our bootstrapping approach by an example. To begin, we generate a multivariate time series according to an instance of eq.~\eqref{eq1} and plot its $N=4$ components in \textbf{Figure \ref{flowchart}A}. We then estimate the causal dependencies from these data by applying a time series causal discovery algorithm, here PCMCI+, which outputs the graph in \textbf{Figure \ref{flowchart}E}. This estimated graph (1) deviates from the ground truth causal graph in \textbf{Figure \ref{flowchart}D} due to finite-sample effects and (2) does not entail any information about the associated uncertainty respectively confidence of output links. Bootstrapping helps to address both of these problems. To this end, we randomly generate $B$ bootstrap datasets from the original data, here $B=100$, using the sampling approach explained in Sec.~\ref{sec:sampling-approach}, where $B$ is the number of chosen bootstrap realizations. Next, we individually apply the to-be-bootstrapped algorithm (here PCMCI+) to all $B$ bootstrap datasets, thus giving rise to $B$ estimated graphs, see \textbf{Figure \ref{flowchart}B}. We now use this ensemble of estimated graphs in two ways. 

First, for bagging, we aggregate the entire ensemble of graphs into a single graph by majority voting (see aggregation prescription in Sec.~\ref{sec:aggregation-approach}), which here outputs the graph in \textbf{Figure \ref{flowchart}C}. The intuition is the following: PCMCI+ is asymptotically consistent but can make errors on finite samples (e.g., false positives and false negatives regarding adjacencies and orientations). Random finite-sample effects tend to cancel out in the aggregated graph, thus making it on average a more accurate estimate than the single estimate on the original data. Specifically, we expect that false links tend to appear in the minority of bootstrap graphs, such that there are fewer false positives in the aggregated graph.

Second, for uncertainty quantification, we count how often each individual edge type (no link, $\tailhead$, $\headtail$, $\oo$, or  $\xx$) of a specific link between each pair $(X^i_{t-\tau}, X^j_t)$ of the aggregated graph appears in the ensemble of graphs and employ the according frequencies as proxies for the confidence in the respective edges. For example, if an edge type for a specific link occurs in $80\%$ of the ensemble of graphs, we have stronger confidence in that link than in a different link where the majority edge type occurs only among $45\%$ of the ensemble of graphs. In \textbf{Figure \ref{flowchart}C}, we visualize the frequencies by the thickness of edges, and we further motivate and discuss this approach to uncertainty quantification in Sec.~\ref{sec:confidence-approach}.

\subsection{Temporal-dependencies preserving sampling approach}\label{sec:sampling-approach}
Our sampling approach is suitable for all algorithms that, as PCMCI+, internally use a moving-window scheme to generate the sets of samples upon which they further operate.

Specifically, PCMCI+ tests marginal and conditional independencies $X \perp\!\!\!\perp Y ~|~ Z$ where $X$, $Y$, and the potentially empty $Z$ can contain variables at different time steps; for example, $X = \{X^1_{t-1}\}$, $Y = \{X^2_t\}$ and $Z = \{X^3_{t-2}, X^2_{t-1}\}$. To create samples for these tests, the algorithm makes a stationarity assumption and employs the following moving window approach: Let $\mathbf{D} = \{w_s ~|~ s \in \mathcal{I}_{\mathbf{D}}\}$ with $\mathcal{I}_{\mathbf{D}} = \{0, 1, \ldots, T-2, T-1\}$ be the time series dataset where $w_s = (x^1_s, x^2_s, \ldots, x^N_s)$ are the measured values at time $s$. Then, for testing $X \perp\!\!\!\perp Y ~|~ Z$ with $X = \{X^1_{t-1}\}$, $Y = \{X^2_t\}$ and $Z = \{X^3_{t-2}, X^2_{t-1}\}$ the algorithm uses the set of samples $\mathbf{S}^{(X,Y,Z)} = \{v^{(X,Y,Z)}_s ~|~ s \in \mathcal{I}_{\mathbf{S}}\}$ with index set $\mathcal{I}_{\mathbf{S}} = \{2\cdot\tau_{\max}, 2\cdot\tau_{\max}+1, \ldots, T-2, T-1\}$ and samples $v^{(X,Y,Z)}_s = (x^1_{s-1}, x^2_s, x^3_{s-2}, x^2_{s-1})$. The choice that $\mathcal{I}_{\mathbf{S}}$ starts with $2 \cdot \tau_{\max}$ instead of $0$ is specific to PCMCI+ and ensures that all independence tests employ the same number of samples, but this choice is irrelevant to our sampling approach described here.

To bootstrap in a non-temporal setting, one would draw bootstrap datasets $\mathbf{D}^\ast$ from the original dataset $\mathbf{D}$ by letting $\mathbf{D}^\ast = \{w_s ~|~ s \in \mathcal{I}^{\ast}_{\mathbf{D}}\}$ where $\mathcal{I}^{\ast}_{\mathbf{D}}$ is sampled with replacement from $\mathcal{I}_{\mathbf{D}}$. Using this procedure in the temporal setting and then applying the moving window approach to $\mathbf{D}^\ast$ would, however, combine values from possibly far away time steps into single samples. Such combinations of values do not reflect the actual temporal dependencies, destroy lag relationships, and, hence, spoil the downstream use of the set of samples (for PCMCI+, spoil the independence testing). Thus, we instead sample with replacement the index set $\mathcal{I}^{\ast}_{\mathbf{S}}$ from $\mathcal{I}_{\mathbf{S}}$ and let $\mathbf{S}^{\ast, (X, Y, Z)} = \{v_s^{(X, Y, Z)} ~|~ s \in \mathcal{I}^{\ast}_{\mathbf{S}}\}$. Here, $v_s^{(X, Y, Z)}$ is exactly as in the previous paragraph and combines values from different time steps in the correct way.
For each bootstrap realization, we run the to-be-bootstrapped causal discovery algorithm in combination with the temporal-dependencies preserving sampling approach. That is, the causal discovery algorithm internally operates with sets of samples which are resampled with the aforementioned approach. We stress that, as in the standard bootstrap, resampling happens only once per bootstrap realization. For each of the $b$ boostrap realizations (where $1 \leq b \leq B$), the set $\mathcal{I}_{\mathbf{S}}^{\ast, (b)}$ is drawn once from $\mathcal{I}_{\mathbf{S}}$, and then the sets $\mathbf{S}^{\ast, (X, Y, Z)} = \{v_s^{(X, Y, Z)} ~|~ s \in \mathcal{I}^{\ast,(b)}_{\mathbf{S}}\}$ are used \emph{for all} (conditional) independence tests that are called in the $b$-th boostrap realization. The to-be-bootstrapped algorithm outputs a causal graph for each bootstrap realization, thus generating an ensemble $\mathcal{C} = \{\mathcal{G}_1,\dots,\mathcal{G}_B\}$ of $B$ causal graphs as an intermediate result.

\subsection{Aggregation by edge-wise majority vote}\label{sec:aggregation-approach}

The prescription for aggregating the bootstrap ensemble $\mathcal{C} = \{\mathcal{G}_1,\dots,\mathcal{G}_B\}$ of estimated graphs to a single final estimated graph $\mathcal{G}_{bagged}$ is independent of the sampling approach. While a variety of choices seems possible, in this work we choose to employ the following strategy: \emph{majority voting}.

In Appendix \ref{appendix:alg}, \textbf{Alg. \ref{alg:majority_voting}} summarizes the aggregation of the $B$ causal graphs \(\mathcal{C} = \{\mathcal{G}_1,\dots,\mathcal{G}_B\}\) to a single final output graph $\mathcal{G}_{bagged}$. For each ordered pair of distinct vertices $(X^i_{t-\tau}, X^j_t)$, we first run through the ensemble of $\mathcal{C}$ and record the relative frequency of each of the possible edge types. For PCMCI+, the possible edge types are \emph{no edge} and $\tailhead$ for lagged pairs ($\tau > 0$), and for contemporaneous pairs ($\tau = 0$) there additionally are the edge types $\headtail$, $\oo$, and $\xx$. Here, the edges $X^i_{t} \oo X^j_t$ and $X^i_{t} \xx X^j_t$ indicate inconclusiveness respectively conflicting information about the direction of the link.\footnote{Conflicting information can arise due to a violation of assumptions or incorrect results of the independence tests.} In $\mathcal{G}_{bagged}$, we then connect $X^i_{t-\tau}$ and $X^j_t$ by an edge of the type with the highest frequency (where connection by \emph{no edge} means that the vertices are not, actually, connected by an edge). To resolve potential ties, we employ the preference order \emph{no edge}, $\xx$, $\oo$, and, at the same preference level, $\tailhead$ and $\headtail$ (from highest to lowest). In case of a tie between only $\rightarrow$ and $\leftarrow$, we resolve to a conflict-indicating link $\xx$. This choice of preference order is conservative in the sense that, among edge types other than \emph{no edge}, it prefers edge types that convey less conclusive claims.

The adaptation of this aggregation strategy to other sets of possible edge types and other tie-resolving strategies is straightforward.
For example, we have also explored an \emph{alternative} aggregation strategy. In the first step of this alternative approach, the orientation of edges is ignored, and the focus is only on determining the adjacency of each pair of vertices. This is done through majority voting between \emph{no edge} and all other edge types. In the second step, the adjacencies identified in the first step are oriented based on majority voting. This alternative approach ensures that \emph{no edge} can only be voted on if it appears in more than half of the bootstrap ensemble of graphs.

The key advantage of aggregating at the level of individual edges is the simplicity of implementation and interpretation. However, this way of aggregating does not in general preserve acyclicity and, more generally speaking, graphical properties that the to-be-bootstrapped algorithms might presuppose. One might view this non-preservation of graphical properties as a disadvantage. Alternatively, one might view this property as a useful feature. The presence of cycles points the method users to a large uncertainty in the respective parts of the graph and to a potential violation of the respective graphical assumption, thus advising them to interpret the results with great care.
It is not hard to come up with alternative aggregation methods that preserve acyclicity. For example, one could first aggregate the graphs as we currently do and then, for each cycle in the aggregated graph, remove the edge that appears with the lowest frequency in the ensemble $\mathcal{C}$. We leave the exploration of this and other alternative aggregation methods, for example using current techniques that minimize a modified structural Hamming distance of the aggregated graph to the entire set of graphs \citep{Wang2014LearningDA}, to future research.

\subsection{Edge frequencies as confidence scores}\label{sec:confidence-approach}
Assuming stationarity, the original time series dataset $\mathbf{D}$ is a (non-\emph{iid}) sample from the stationary distribution of the stochastic process defined by the respective instance eq.~\eqref{eq1}. The sample set $\mathbf{S}^{(X, Y, Z)}$ defined in Sec.~\ref{sec:sampling-approach} is, thus, a (non-\emph{iid}) sample from a $t$-independent distribution $F^{(X, Y, Z)}$. This sample set defines an empirical distribution $\hat{F}^{(X,Y,Z)}_T$ that approximates $F^{(X, Y, Z)}$, and by the design of our sampling approach the bootstrap sample set $\mathbf{S}^{\ast, (X, Y, Z)}$ is a sample from $\hat{F}^{(X,Y,Z)}_T$.

Apart from autodependencies, we are hence in the standard bootstrap setting and can use the bootstrap to approximate uncertainties with respect to $F^{(X, Y, Z)}$. Specifically, we can quantify the uncertainty respectively confidence in a given edge $e$ in the aggregated graph $\mathcal{G}_{bagged}$ by the frequency $r_e^{boot}$ with which this edge appears in the bootstrap ensemble $\mathcal{C}$ of estimated graphs. The precise meaning of this measure of uncertainty is as follows: As $T \to \infty$ and $B \to \infty$, we expect that $r_e^{boot}$ converges to the frequency $r_e$ with which the same edge $e$ appears in the hypothetical ensemble of graphs obtained by repeatedly applying the to-be-boostraped method on time series datasets $\mathbf{D}_1, \mathbf{D}_2, \ldots$ that are generated \emph{from the model} (as opposed to sampling from $\mathbf{D}$). Thus, a higher $r_e^{boot}$ means a higher confidence in the edge $e$. In our visualizations of aggregated graphs, we represent $r_e^{boot}$ by the edge's thickness, where edges with higher $r_e^{boot}$ are thicker. 

The advantages of this edge-wise uncertainty measure are (1) its simplicity and (2) that it conforms well with the edge-wise aggregation of the bootstrap ensemble $\mathcal{C}$ of graphs to $\mathcal{G}_{bagged}$. However, in the bootstrap ensemble, the presence, absence and orientation of an edge might be correlated with the presence, absence and orientation of other edges. Since our aggregation scheme operates \emph{edge-wise}, that is, on the level of individual edges, the aggregation does not take into these potential dependencies between edges. As a result, information about such dependencies is lost during the edge-wise aggregation process.

\section{Numerical experiments}
\label{section_num_exp}
\subsection{Evaluation of Bagged-PCMCI+ performance on synthetic data} \label{subsection:eval}

The numerical experiments model several typical challenges in time series causal discovery~\citep{Runge2019InferringCF}: contemporaneous and time-lagged causal dependencies, nonlinearity, non-Gaussian noise distribution, strong autocorrelation, large numbers of variables and considered time lags. For a better comparison to PCMCI+, we use a similar setup to the numerical experiments presented in \citet{Runge2020DiscoveringCA}.
The synthetic data is generated according to the following additive model:
\begin{equation}
\label{eq:exp}
X^j_t = a_j X_{t-1}^{j} + \sum_{i}^{} c_i f_i(X^i_{t-\tau_i}) + \eta_t^j 
\quad \forall j \in \{1, \dots, N\} \end{equation} 
Autocorrelation coefficients \begin{math} a_j\end{math} are uniformly drawn from [max\begin{math}(0, a - 0.3),a] \end{math}. The values of the autocorrelation \(a\) and of other parameters of the data-generating model are indicated in the header of Figures \ref{exp_prcurve}, \ref{exp_aucplot}, \ref{truevbootfreq}, and Figures in the Appendix.
The noise terms \(\eta^j\) are independent and identically distributed zero-mean Gaussians
\(\mathcal{N}(0,\sigma^2)\) with standard deviation $\sigma$ drawn from \([0.5, 2]\) or Weibull (scale parameter 2) depending on the setup.
In addition to autodependency links, for each model \(L = \lfloor1.5\cdot N\rfloor\) (except when \(N = 2\) where we set \(L = 1\)) cross-links are chosen with linear functional dependencies \(f_i(x) = x\) or with nonlinear functional dependencies $f_i(x) = x + 5 x^2 e^{-x^2/20}$ depending on the setup.
Coefficients \(c_i\) are drawn uniformly from \(\pm[0.1, 0.5]\). \(30\%\) of the links are contemporaneous (\(\tau_i = 0\)) and the remaining 70\% are lagged links with \(\tau_i\) drawn from \(\{1, \dots, 5\}\).
Only stationary models are considered. We have an average cross-in-degree of \(d = 1.5\) for all network sizes (plus an auto-dependency) implying that models become sparser for larger N. We consider several model setups: linear with Gaussian noise, linear with mixed noise (50\% Gaussian and 50\% Weibull), and nonlinear (50\% linear and 50\% nonlinear dependencies) with Gaussian noise. We consider the PCMCI+ algorithm and Bagged($B$)-PCMCI+ where the number of bootstrap realizations \(B\) varies. If no aggregation strategy is mentioned, the summary graph of Bagged($B$)-PCMCI+ is obtained by simple majority voting. To test conditional independence, we use the partial correlation test (ParCorr) for linear experiments or the GPDC test \citep{runge2019detecting} for nonlinear experiments implemented in the tigramite Python package (see Appendix \ref{code_avail}). To show that our bagging approach is general, we also include numerical experiments that, instead of PCMCI+, use a modified PC algorithm adapted to time series and the LPCMCI algorithm \citep{NEURIPS2020_94e70705} as the base algorithm. We chose the adapted PC algorithm because it can be combined with our resampling approach easily, and decided to also experiment with LPCMCI as a state-of-the-art algorithm for time series causal discovery in the presence of contemporaneous edges and hidden confounders. For the experiment with LPCMCI, we randomly choose 70\% from the $N$ variables of each model as observed ($\Tilde{N} = \lceil 0.7 N \rceil$). The rest of the variables are unobserved.

Performance is evaluated with recall (equivalent to True Positive Rate, TPR), precision, and the area under the precision-recall curve (PR-AUC). For adjacencies, precision and recall are distinguished between lagged cross-links (\(i \neq j\)), contemporaneous (\( \tau = 0\)), and autodependency links (\(i = j\)). Due to time order, lagged links (and autodependencies) are automatically oriented. 
Performance of contemporaneous orientation is evaluated with contemporaneous orientation precision, which is measured as the fraction of correctly oriented links (\(\oo\),\(\rightarrow\),\(\leftarrow\)) among all estimated adjacencies, and with recall as the fraction of correct orientations among all true contemporaneous links. False positive rates (FPR) are also shown to evaluate whether the methods control false positives at the chosen significance level $\alpha_{\rm PC}$. Furthermore, the fraction of conflicting links among all detected contemporaneous adjacencies is calculated. Finally, we give the average runtimes that were evaluated on an AMD EPYC 7763 processor. All metrics (and their standard errors) are computed across all estimated graphs from 500 different additive models (and associated realizations) described in Equation \ref{eq:exp} with time series length \(T\).

\begin{figure}[ht]
  \includegraphics[width=0.56\linewidth]{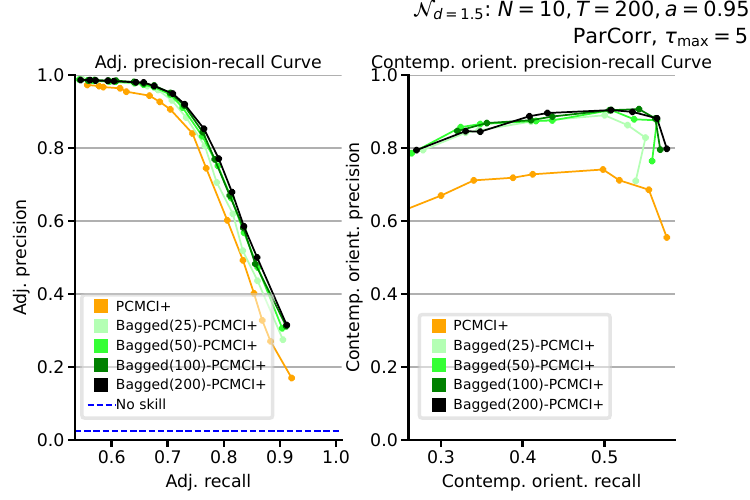}
  \centering
  \caption{Precision-recall curves for adjacencies (left) and contemporaneous orientations (right) obtained by varying the significance level \(\alpha_{PC}\) in PCMCI+ and Bagged-PCMCI+ for the model setup as shown in the top right. Results are shown for PCMCI+ (orange line) and Bagged-PCMCI+ with different numbers of bootstrap replicas $B$ (lines with different shades of green).}
  \label{exp_prcurve}
\end{figure}

\begin{figure}[ht]
  \includegraphics[width=0.46\linewidth]{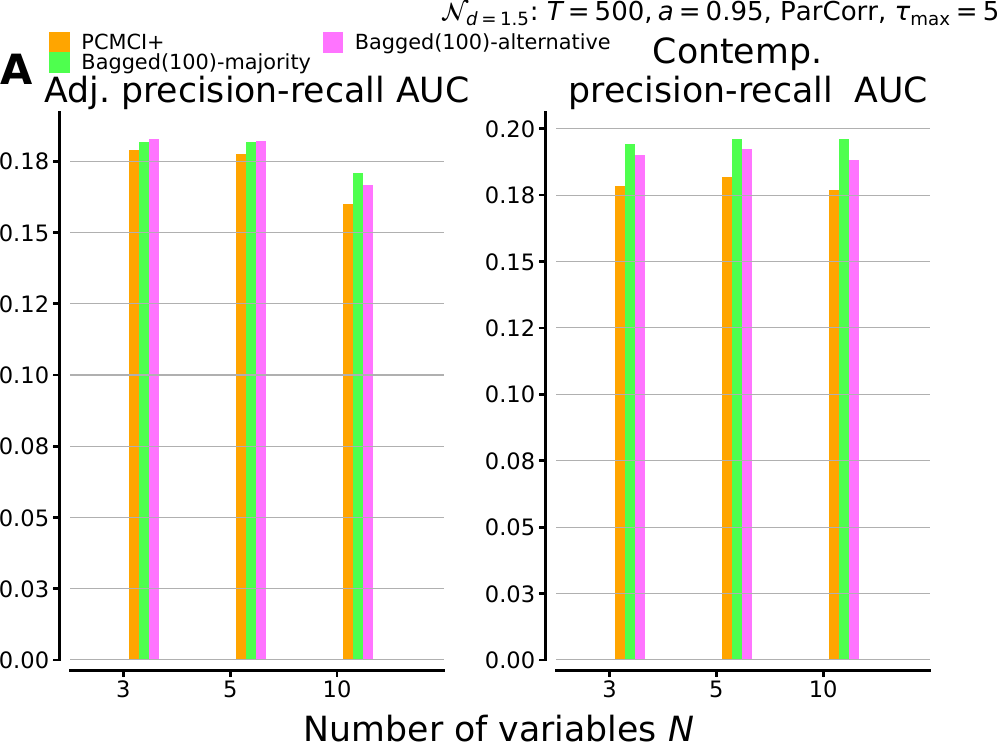}
  \includegraphics[width=0.46\linewidth]{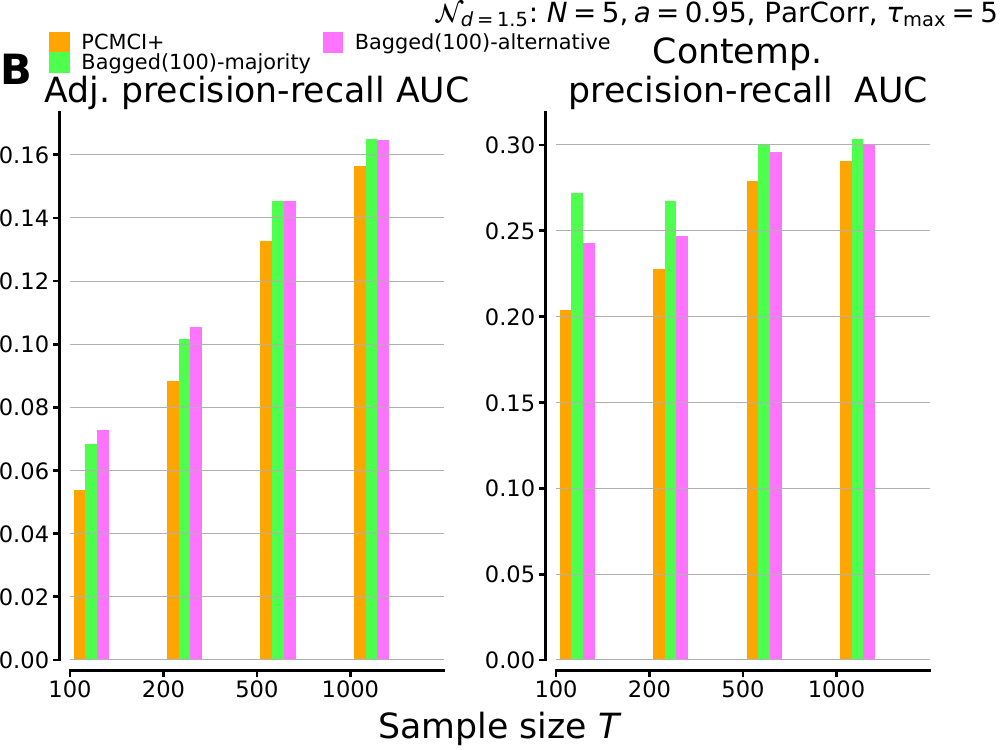} \\
  \includegraphics[width=0.46\linewidth]{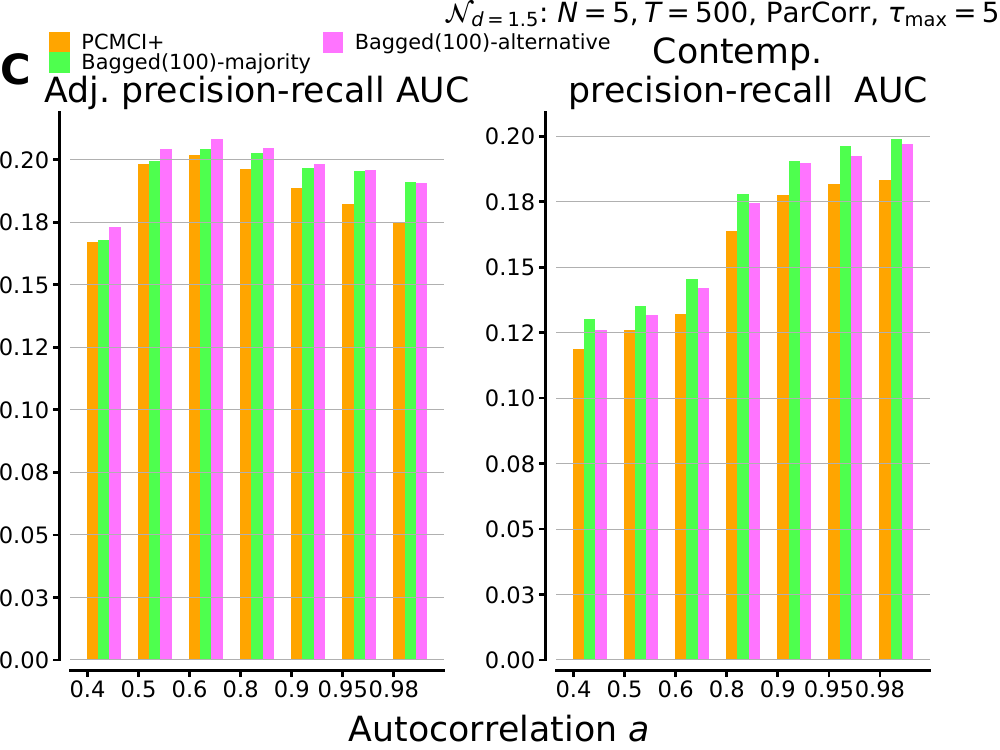}
  \includegraphics[width=0.46\linewidth]{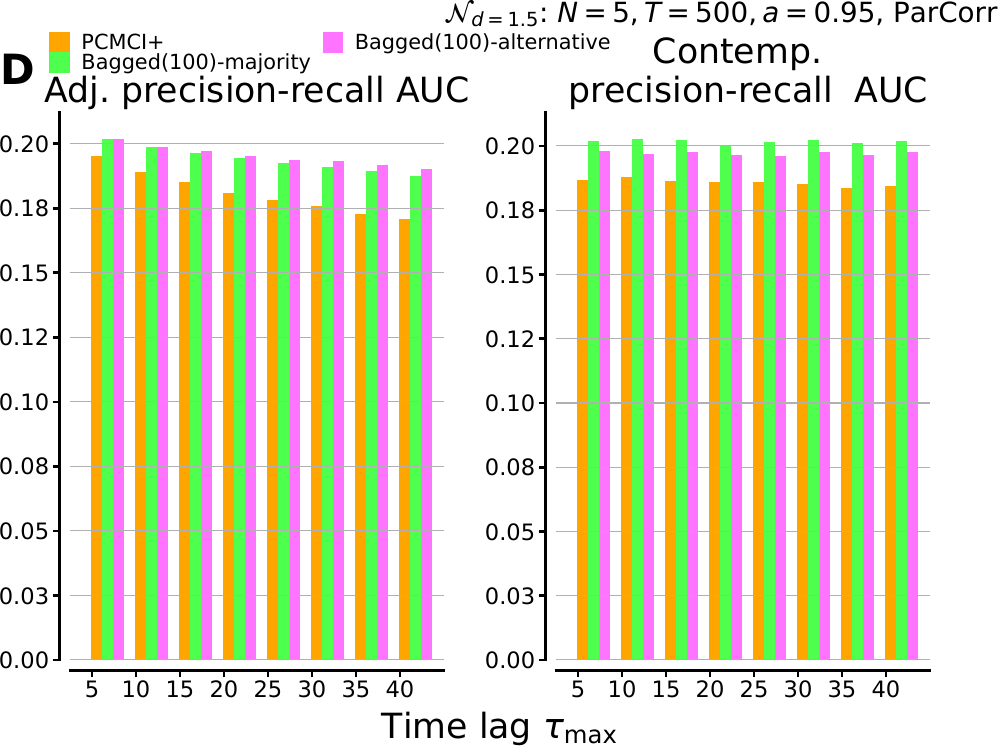} \\
  \centering
  \caption{PR-AUC of adjacencies and contemporaneous orientations for the linear and Gaussian noise setup for a varying: (\textbf{A}) number of variables $N$, (\textbf{B}) sample size $T$, (\textbf{C}) autocorrelation $a$, (\textbf{D}) PCMCI+ maximum time lag $\tau_{\max}$. The model setup is shown at the top right. The Precision-recall curves are obtained by varying the significance level \(\alpha_{PC}\) of PCMCI+. PR-AUC is the area under these curves. The results for PCMCI+ (orange bars), Bagged-PCMCI+ with $B=100$ bootstrap replicas for the majority voting aggregation strategy (green bars) and the alternative aggregation strategy (pink bars) are shown.}
  \label{exp_aucplot}
\end{figure}

\textbf{Figure~\ref{exp_prcurve}} depicts \textbf{precision-recall curves} for adjacencies as well as contemporaneous orientations obtained by varying the hyperparameter \(\alpha_{PC}\) for a model setup that stands exemplary for others (see Appendix \ref{appendix:pcmci+}). The precision-recall values of Bagged-PCMCI+ are systematically higher than those of PCMCI+, regarding adjacencies and even more pronounced for contemporaneous orientations. Moreover, larger numbers of bootstrap replicates \(B\) result in enhanced performance, but we observe no strong differences between $B=50$ and $B=200$. More precision-recall curves are shown in Appendix \ref{appendix:pcmci+} for different model parameters (autocorrelation, number of variables, and time sample size). In \textbf{Figure \ref{exp_aucplot}}, we plot the area under the precision-recall curves for varying model parameters. Bagged-PCMCI+ has systematically \textbf{higher PR-AUC} than PCMCI+, indicating that Bagged-PCMCI+ outperforms PCMCI in terms of precision and recall. Noticeably, we found that Bagged-PCMCI+ outperforms PCMCI+ more clearly in statistically more challenging regimes with higher autocorrelation \(a\), larger number of variables \(N\), and smaller time sample sizes \(T\). We also compare the PR-AUC of Bagged(100)-PCMCI+ with the simple majority voting strategy to the alternative aggregation strategy introduced in Section \ref{sec:aggregation-approach}. The alternative aggregation strategy generally leads to improvement in terms of the PR-AUC for the adjacencies, whereas the simple majority voting strategy is superior regarding contemporaneous orientations.

In the Appendix \ref{appendix:pcmci+}, \textbf{Figure~\ref{exp_alphapc}} gives further details for the linear Gaussian setup for varying significance level $\alpha_{PC}$.
The right column of Figure \ref{exp_alphapc} clearly shows that both methods (PCMCI+ and Bagged-PCMCI+) control the FPR below the significance level \(\alpha_{PC}\) (grey line), but Bagged-PCMCI+ consistently exhibits \textbf{lower FPR} compared to PCMCI+ for all types of links (lagged, contemporaneous, auto links) and across significance levels \(\alpha_{PC}\) ranging from 0.001 to 0.1. Thus, treating the significance level as a hyperparameter, one can use a higher level $\alpha_{\rm PC}$ for Bagged-PCMCI+ than for PCMCI+ while still controlling the FPR below \(\alpha_{PC}\).
There appear to be slightly fewer conflicts for Bagged-PCMCI+. Runtimes are, as expected, higher for Bagged-PCMCI+, but these can be reduced by embarrassingly parallelization.
In the Appendix \ref{appendix:pcmci+}, we provide additional figures for varying one of autocorrelation \(a\), number of variables \(N\), time sample size \(T\), and maximum time lag \(\tau_{\max}\) with \(\alpha_{PC} = 0.01\). We found that Bagged-PCMCI+ seems robust to large maximum time lags \(\tau_{\max}\) (even when \(\tau_{\max}\) is much larger than the true maximum time lag of 5) for the studied sample size \(T = 500\). We have observed similar improvements of Bagged-PCMCI+ in terms of precision and recall compared to PCMCI+ in the nonlinear and mixed noise experiments (see \textbf{Appendix \ref{nonlinear_mixed}}).

Finally, we have combined our bagging approach with a modified PC algorithm adapted to time series and LPCMCI. We provide results for these experiments in the \textbf{Appendix \ref{section:pc}} and \textbf{Appendix \ref{section:lpcmci}} respectively. Similar to PCMCI+, we have found that Bagged-PC enhances its base algorithm PC in terms of precision and recall. For Bagged-LPCMCI, we have identified improvements in terms of contemporaneous orientation precision and recall relative to LPCMCI.

\subsection{Evaluation of Bagged-PCMCI+ confidence measure}

We conduct numerical experiments to assess the ability of Bagged-PCMCI+ to determine a confidence degree for links in the output graph. Ideally, we would like the Bagged-PCMCI+ link frequency obtained on a single data sample to approximate closely the frequency of links along graphs obtained independently by PCMCI+ on an infinite number of data samples, which we call reference link frequencies. In practice, it is only possible to approximate the reference link frequencies by using a large but limited number of data samples. Here, we design two experiments to evaluate the ability of the proposed confidence measures to approximate the reference link frequencies.

\begin{figure}[ht]
  \centering
  \includegraphics[width=0.8\linewidth]{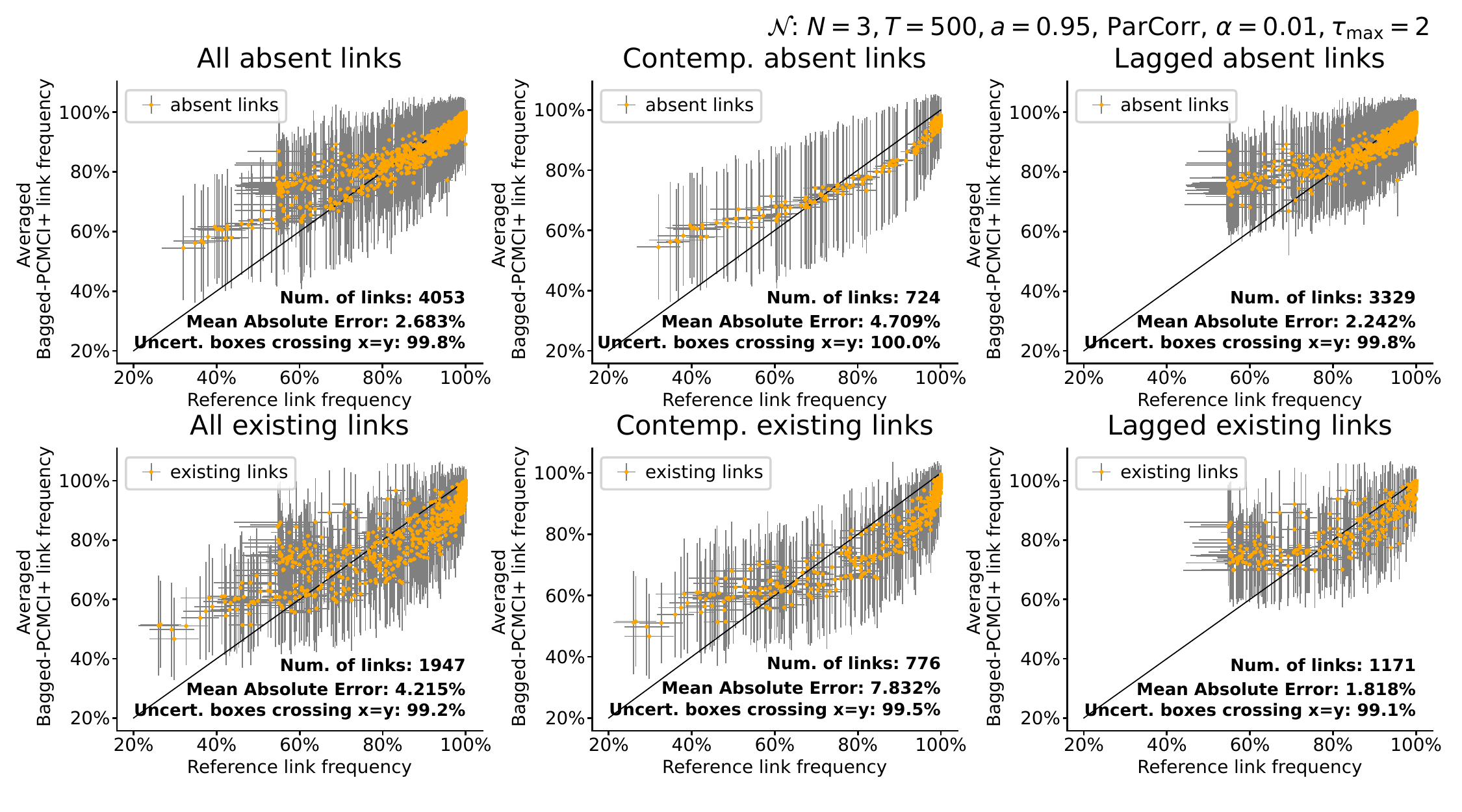}
  \caption{Evaluation of the proposed confidence measures for a linear Gaussian setup with parameters indicated at the top right. $y$-values (averaged confidence measure) and $x$-values (reference) are as explained in the text. Grey bars indicate the one standard deviation error around the estimated value.
  }
  \label{truevbootfreq}
\end{figure}

In the \textbf{first experiment}, we generate 250 different additive models (see Equation \ref{eq:exp}). For each of these models, we generate \(D=100\) independent data samples with the same additive model. That is, only the noise terms change across the samples. For each of these 100 samples, causal graphs are estimated independently using PCMCI+ and Bagged-PCMCI+. For each edge, we estimate its \emph{reference link frequency} by calculating the frequency of the most frequent link types across the PCMCI+ ensemble of 100 graphs. We use the mean of our proposed confidence measure (i.e. the mean of the Bagged-PCMCI+ link frequencies along the 100 Bagged-PCMCI+ graphs) to reduce the amount of noise in the estimation. We also calculate the standard deviation of the Bagged-PCMCI+ link frequencies along the 100 Bagged-PCMCI+ graphs to estimate its uncertainty. If effective, we expect the Bagged-PCMCI+ confidence measure to approximately follow the estimated reference link frequency. In \textbf{Figure \ref{truevbootfreq}}, we show the results of this first experiment for model parameters and method parameters indicated at the top right and for \(B=1000\). We plot averaged confidence measures ($y$-axis) against the reference link frequencies ($x$-axis) for different causal dependencies (lagged, contemporaneous, or all) and for existing/absent links of the ground truth graphs. Figure \ref{truevbootfreq} shows that the Bagged-PCMCI+ link frequency (confidence measure) overall tends to follow the reference link frequency. We can notice a bias for low reference link frequencies as Bagged-PCMCI+ tends to overestimate the reference link frequencies between 40\% to 60\%. This bias seems consistent across different types of causal dependencies. The one standard deviation error bars give indications of the uncertainties in the confidence measure and the estimated reference link frequency. When taking both uncertainties into account, the error bars cross the expected \(x=y\) diagonal line more than 99\% of the estimated link frequencies. This high percentage is a good indication that the Bagged-PCMCI+ confidence measure approximates the reference link frequency of PCMCI+ reasonably well.

In the \textbf{second experiment}, we quantify the average absolute error of the Bagged-PCMCI+ link frequency relative to the reference link frequency for different causal dependencies (lagged, contemporaneous, or all) and for existing links and absent links of the ground truth model. For each additive model, the reference link frequency is calculated as the frequency of the most recurrent link type of PCMCI+ graphs over \(D=5000\) independent samples. Here, the Bagged-PCMCI+ link frequency is a confidence measure for the first data sample of each additive model (the confidence is, as compared to the first experiment, not averaged). We calculate the mean absolute error between the confidence measures and the estimated reference link frequencies. We generate a total of 500 different additive models and average the mean absolute link frequency errors. To study the effect of the number of bootstrap realizations \(B\) on the error, we vary \(B\) from 25 to 2000. We present the results of the mean absolute link frequency errors in \textbf{Fig \ref{truevbootfreq2}}. For lagged links and all links, the mean absolute errors are slightly below 3\% which seems relatively low. For the contemporaneous links, the 7\% errors demonstrate that the estimation of the contemporaneous links is a more difficult task for the current method. It is also not clear if the mean absolute error converges to zero if \(B\) goes to infinity. We observe that a larger \(B\) leads to a lower mean absolute error of Bagged-PCMCI+. This confirms that increasing \(B\) enhances the performance of the Bagged-PCMCI+. The mean absolute frequency error reduces by about 10\% when increasing \(B\) from 25 to 500. 
Unfortunately, without parallelization, the increase in performance comes at the cost of a 20-fold increase in runtime. This is why we recommend using a number of bootstrap realizations which is adapted for the application and available computing resources, but preferably larger or equal to 100.

\section{Conclusion} \label{section:ccl}

In this paper, we have made \textbf{two major contributions} to time series causal discovery. First, we propose a \textbf{bootstrap aggregation} by majority voting that can be combined with any time series causal discovery algorithm. Here, we combine our bootstrap aggregation approach with the state-of-the-art time series causal discovery PCMCI+ algorithm (referred to as Bagged-PCMCI+). Through extensive numerical experiments, we show that Bagged-PCMCI+ greatly reduces the number of false positives compared to the base PCMCI+ algorithm. In addition, Bagged-PCMCI+ has a higher precision-recall regarding adjacencies and orientations of contemporaneous edges compared to PCMCI+. The same outcomes hold true when combining our bootstrap aggregation approach with a modified version of the PC algorithm or with the LPCMCI algorithm, suggesting that our numerical results are consistent and generalizable. Our second contribution is a \textbf{confidence measure for links} in a time series graph, which is calculated as the link frequencies along the graphs learned on bootstrap replicates. Numerical experiments show that the proposed method gives a pertinent confidence measure for links of the output graph.
The \textbf{main strengths} of our method are that it can be coupled with a wide range of time series causal discovery algorithms, and it can be of substantial use in many real-world applications, especially for orienting contemporaneous causal links or when confidence measures for links are desired. The \textbf{main weaknesses} of our method so far are its higher computational cost and longer runtime. One solution to decrease runtime is to embarrassingly parallelize the bootstrap process. In addition, the current method of aggregating through majority voting has a limitation. It can lead to cyclic graphs that are not always desirable. Therefore, exploring alternative methods of aggregation will constitute a crucial step for future research.

\acks{DKRZ provided computational resources (grant no.
1083). K.D. and V.E have received funding provided by the European Research Council (ERC) Synergy Grant “Understanding and Modelling the Earth System with Machine Learning (USMILE)” under the Horizon 2020 research and innovation programme (Grant agreement No. 855187). J.R. has received funding from the European Research Council (ERC) Starting Grant CausalEarth under the European Union’s Horizon 2020 research and innovation program (Grant Agreement No. 948112). We thank Tom Hochsprung for his helpful comments.}

\bibliography{refs}

\appendix

    
\section{Code availability}\label{code_avail}

Code for the bootstrap aggregation of PCMCI+, LPCMCI, and other variants as well as code for the conditional independence tests ParCorr and GPDC are provided as part of the tigramite Python package at \url{https://github.com/jakobrunge/tigramite}. Numerical experiments can be reproduced with the code available at \url{https://github.com/EyringMLClimateGroup/debeire24clear\_Bagged\_TimeSeries\_Causality}.

\section{Description of the aggregation by edge-wise majority (pseudo-code)} \label{appendix:alg}
\begin{minipage}{\textwidth}
\renewcommand*\footnoterule{}
\begin{savenotes}
\begin{algorithm}[H]
    \caption{(Aggregation of bootstrap causal graphs by majority voting of edges)}
    \label{alg:majority_voting}
\SetKwInOut{Input}{input}\SetKwInOut{Output}{output}

\Input{Ensemble of bootstrap causal graphs \(\mathcal{C} = \{\mathcal{G}_1,\dots,\mathcal{G}_B\} \)
}

\Output{Aggregated graph \(\mathcal{G}_{bagged}\), the relative  frequency $F$ of link types for each pair of vertices $(X^i_{t-\tau}, X^j_t)$ and link type \(e\) across $\mathcal{C}$}
\BlankLine
Initialize an empty graph \(\mathcal{G}_{bagged}\)\ with the same nodes as in the bootstrap causal graphs\;

Initialize a two-level nested dictionary $F$ for recording relative frequencies of link types\;

\ForAll{pairs $(X^i_{t-\tau}, X^j_t)$ with $0 \leq \tau \leq \tau_{\max}$, $1 \leq i,j \leq N$, and $i < j$ if $\tau=0$}{
\ForAll{possible link types $e$}{
\(n(e) \gets \text{ number of graphs in } \mathcal{C} \text{ in which }(X^i_{t-\tau}, X^j_t) \text{is connected by }e\)\;

\(F[(X^i_{t-\tau}, X^j_t)][e] \gets \tfrac{n(e)}{B}\)\;
}

Link $(X^i_{t-\tau}, X^j_t)$ in $\mathcal{G}_{bagged} \gets \underset{e}{\mathrm{argmax}} \  F[(X^i_{t-\tau}, X^j_t)][e]$ \footnote{Ties are resolved according to the preferred order given in section \ref{sec:aggregation-approach}.} \;
}
\Return{\(\mathcal{G}_{bagged},F\)}
\end{algorithm}
\end{savenotes}
\end{minipage}

\section{Additional Bagged-PCMCI+ experiments}\label{appendix:pcmci+}

\subsection{Further numerical experiments for the linear and Gaussian noise setup}

\begin{figure}[ht]
  \includegraphics[width=0.55\linewidth]{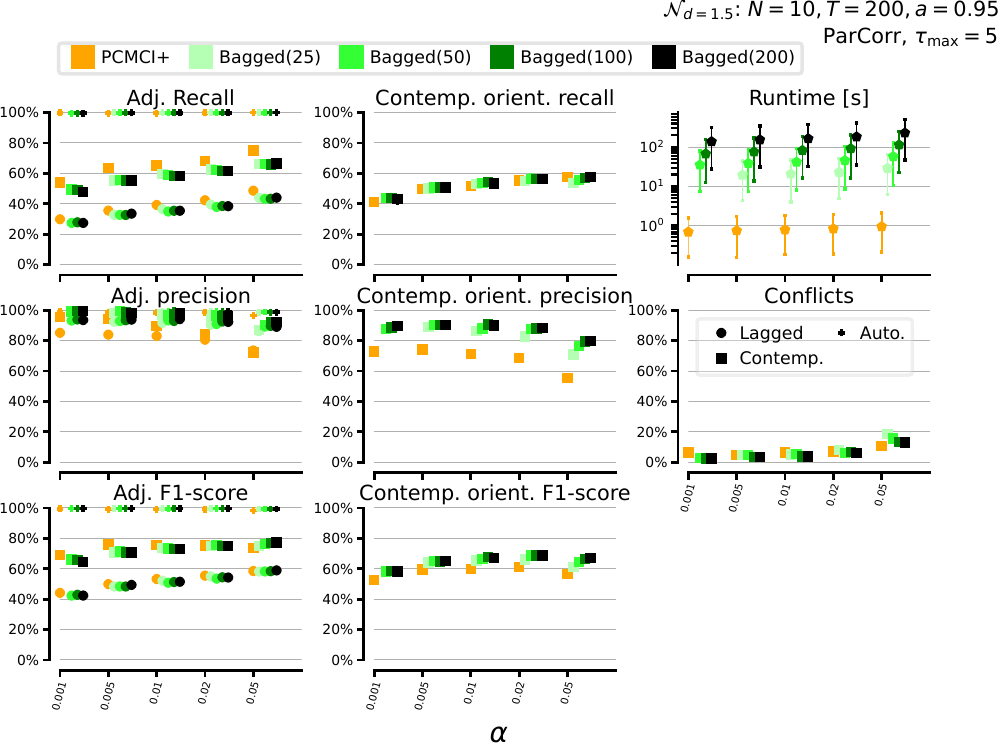}
  \includegraphics[width=0.3\linewidth]{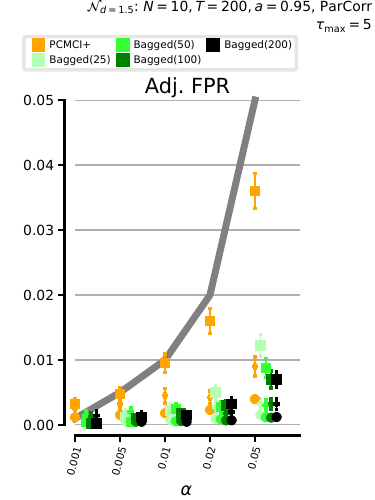}
  \centering
  \caption{Numerical experiments with linear Gaussian setup for varying significance level \(\alpha_{PC}\) of PCMCI+.}
  \label{exp_alphapc}
\end{figure}

\begin{figure}[ht]
  \includegraphics[width=0.7\linewidth]{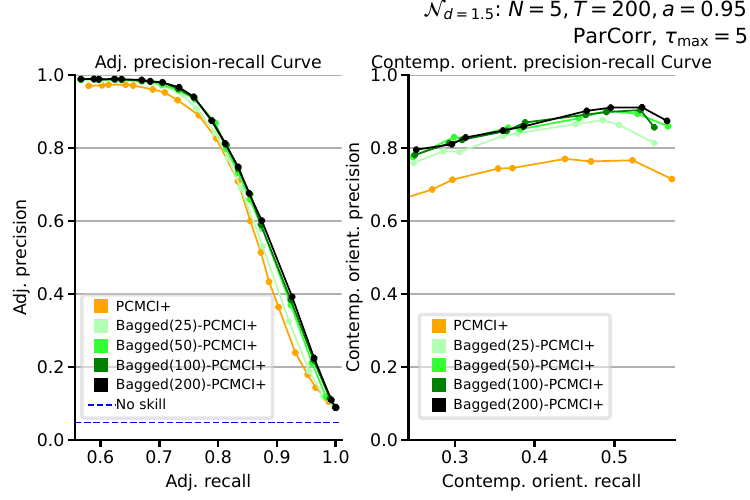}
  \centering
  \caption{Precision-recall curves for adjacencies (left) and contemporaneous orientations (right) obtained by varying the significance level \(\alpha_{PC}\) in PCMCI+ and Bagged-PCMCI+ for the model setup as shown in the header. Results are for PCMCI+ (orange line) and Bagged-PCMCI+ with different numbers of bootstrap replicas $B$ (lines with different shades of green). Here \(N=5\), \(T=200\), and \(a =0.95\).}
  \label{sm_prcurve1}
\end{figure}

\begin{figure}[ht]
  \includegraphics[width=0.605\linewidth]{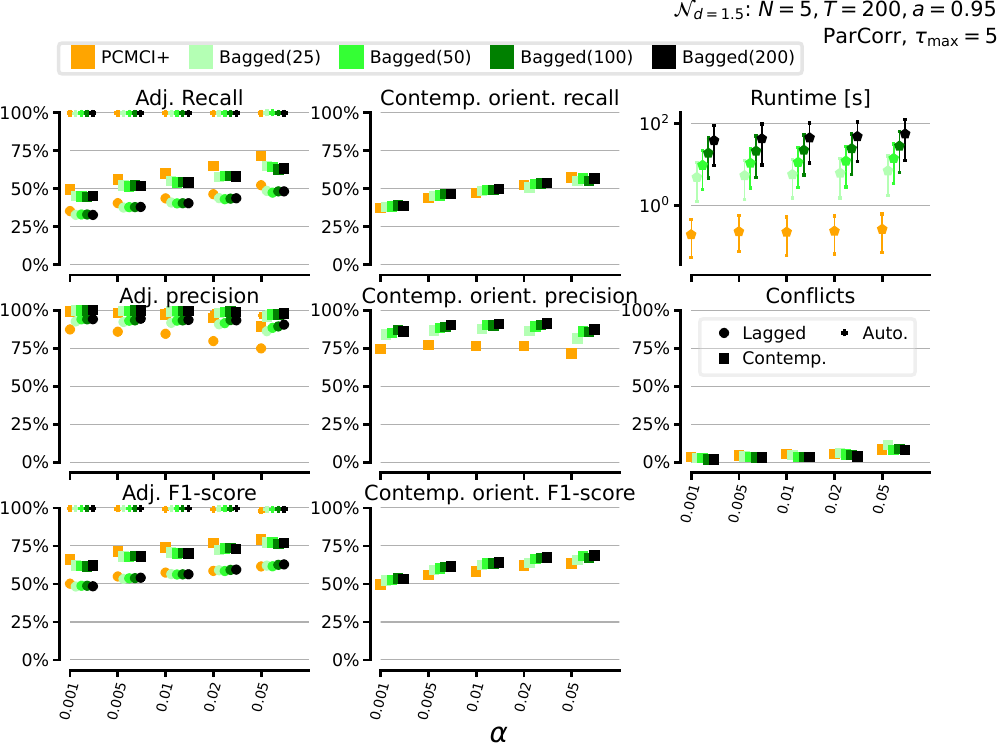}
  \includegraphics[width=0.33\linewidth]{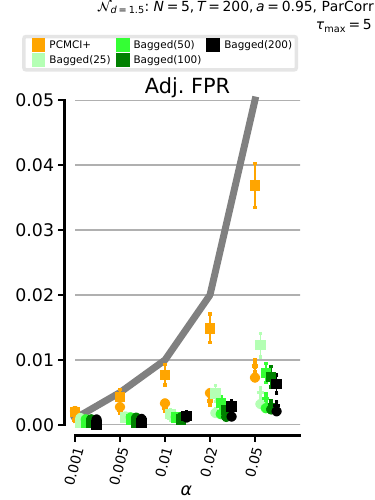}
  \centering
  \caption{Numerical experiments with linear Gaussian setup for varying \(\alpha_{PC}\) of PCMCI+. Here \(N=5\), \(T=200\), and \(a =0.95\).}
  \label{sm_panel1}
\end{figure}

\begin{figure}[ht]
  \includegraphics[width=0.7\linewidth]{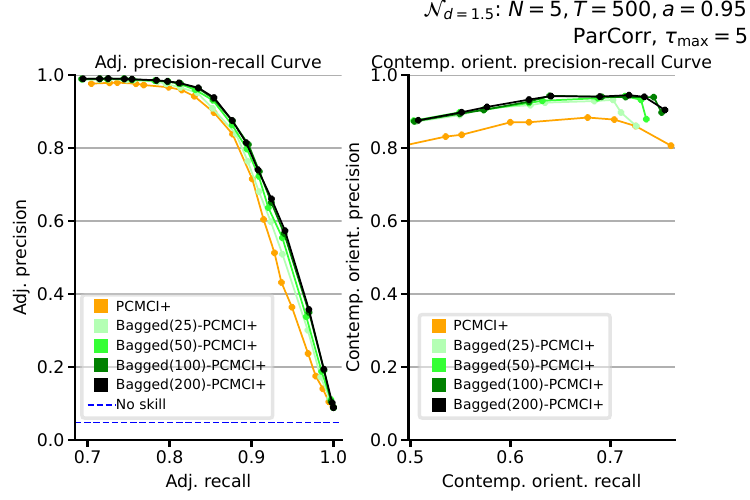}
  \centering
  \caption{Precision-recall curves for adjacencies (left) and contemporaneous orientations (right) obtained by varying the significance level \(\alpha_{PC}\) in PCMCI+ and Bagged-PCMCI+ for the model setup as shown in the header. Results are for PCMCI+ (orange line) and Bagged-PCMCI+ with different numbers of bootstrap replicas $B$ (lines with different shades of green). Here \(N=5\), \(T=500\), and \(a =0.95\).}
  \label{sm_prcurve2}
\end{figure}

\begin{figure}[ht]
  \includegraphics[width=0.605\linewidth]{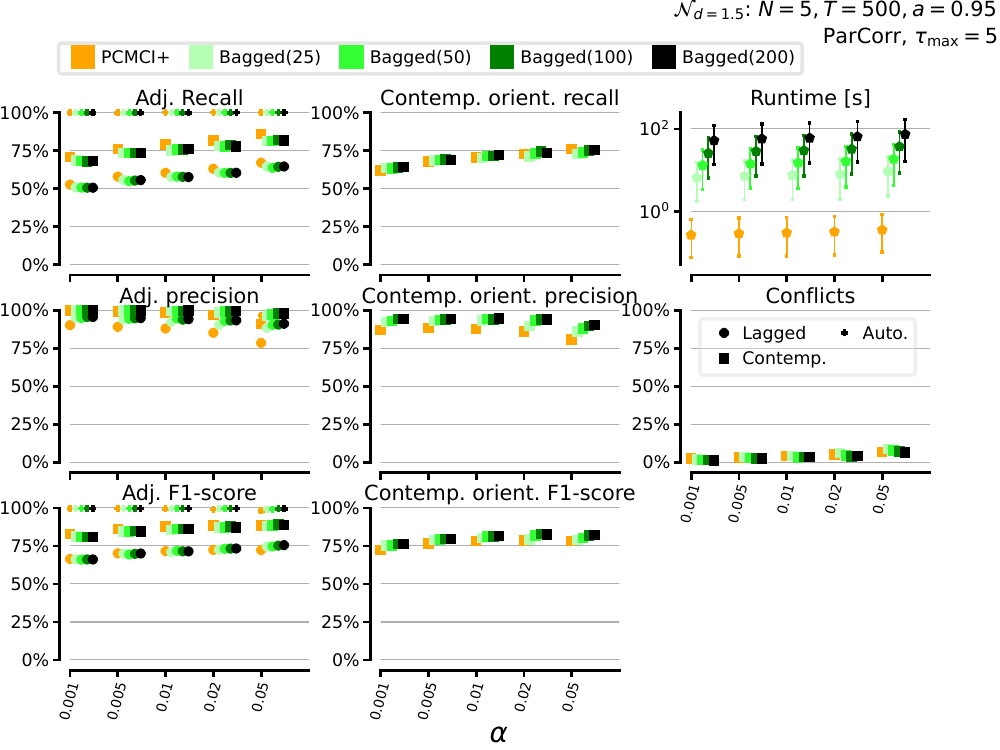}
  \includegraphics[width=0.33\linewidth]{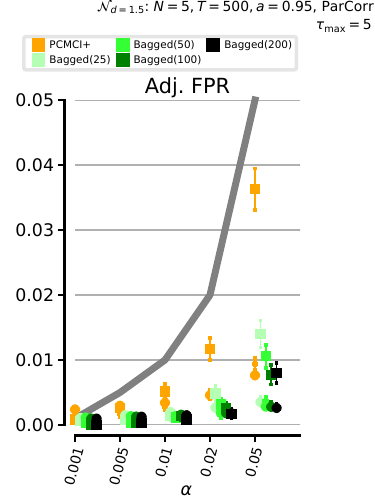}
  \centering
  \caption{Numerical experiments with linear Gaussian setup for varying \(\alpha_{PC}\) of PCMCI+. Here \(N=5\), \(T=500\), and \(a =0.95\).}
  \label{sm_panel2}
\end{figure}

\begin{figure}[ht]
  \includegraphics[width=0.7\linewidth]{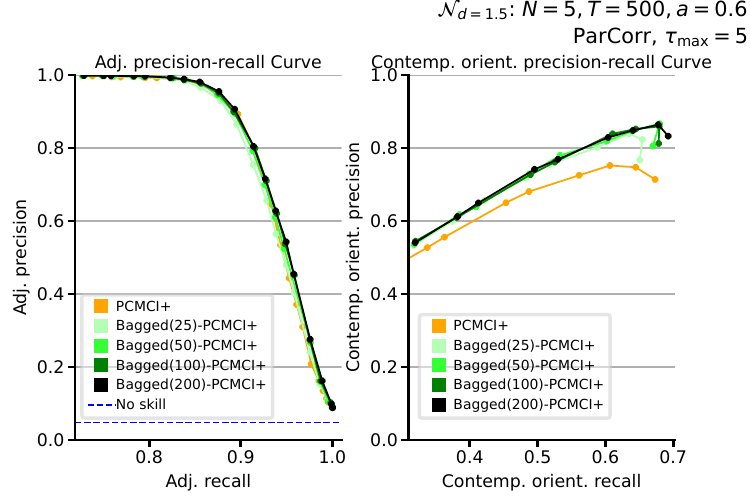}
  \centering
  \caption{Precision-recall curves for adjacencies (left) and contemporaneous orientations (right) obtained by varying the significance level \(\alpha_{PC}\) in PCMCI+ and Bagged-PCMCI+ for the model setup as shown in the header. Results are for PCMCI+ (orange line) and Bagged-PCMCI+ with different numbers of bootstrap replicas $B$ (lines with different shades of green). Here \(N=5\), \(T=500\), and \(a =0.6\).}
  \label{sm_prcurve3}
\end{figure}

\begin{figure}[ht]
  \includegraphics[width=0.605\linewidth]{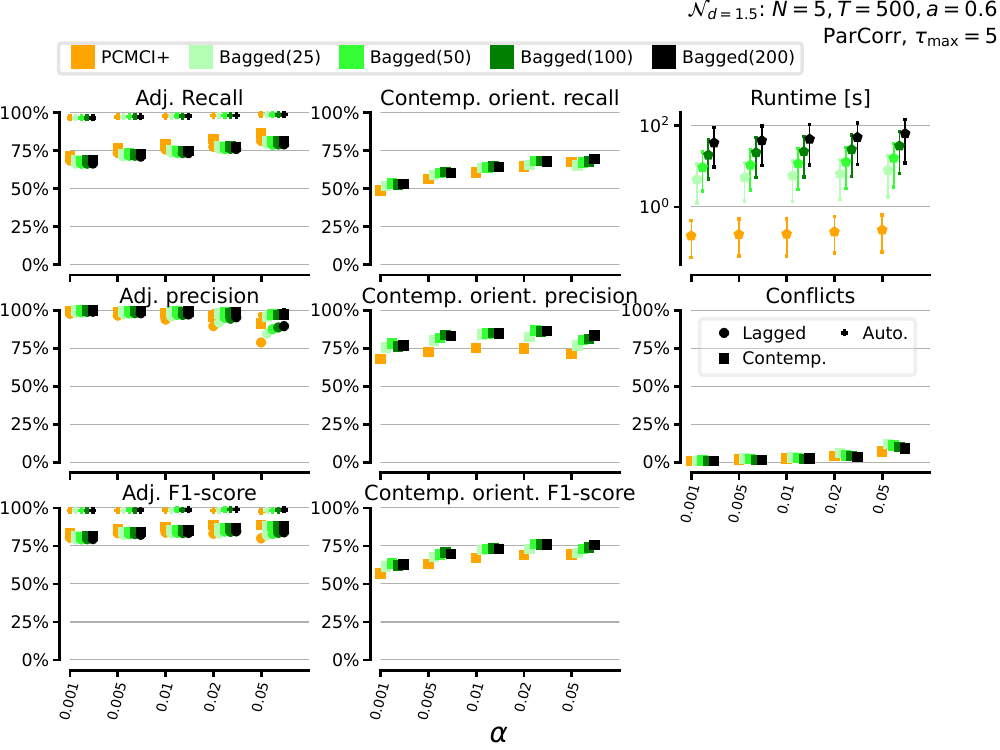}
  \includegraphics[width=0.33\linewidth]{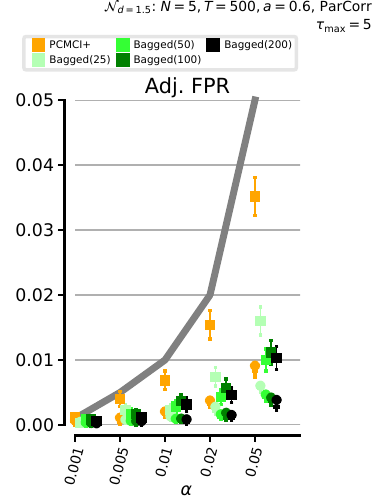}
  \centering
  \caption{Numerical experiments with linear Gaussian setup for varying \(\alpha_{PC}\) of PCMCI+. Here \(N=5\), \(T=500\), and \(a =0.6\).}
  \label{sm_panel3}
\end{figure}

We investigate the impact of model parameters on Bagged-PCMCI+ in the linear and Gaussian noise experiments compared to the model setup of \textbf{Figure \ref{exp_prcurve}} in the main text. Precision-recall curves for additional model setups show the impact of a smaller number of variables of $N=5$ instead of $N=10$ in the main text (\textbf{Figure \ref{sm_prcurve1}}), increased sample size $T$ from $200$ to $500$ for $N=5$ (\textbf{Figure \ref{sm_prcurve2}}), and a decreased autocorrelation coefficient $a$ from $0.95$ to $0.6$ for $T=500$ and $N=5$ (\textbf{Figure \ref{sm_prcurve3}}). For all these model setups we also provide the individual precisions, recalls, F1-scores plots for adjacencies and contemporaneous orientations, as well as the runtimes and number of conflicts for varying $\alpha_{\rm PC}$ (see \textbf{Figures \ref{exp_alphapc}, \ref{sm_panel1}, \ref{sm_panel2}, and \ref{sm_panel3}}). \(F_1\) scores are calculated for adjacencies and contemporaneous orientations as the harmonic mean of precision and recall: \( F_1 = 2 \frac{precision \cdot recall}{precision + recall}\).

Across all these model setups, for a given $\alpha_{\rm PC}$ Bagged-PCMCI+ has similar recall and higher precision as compared to PCMCI+, particularly in orienting contemporaneous links. Moreover, these improvements are stronger in the more challenging settings (high autocorrelation $a$, short time sample size $T$, and high number of variables $N$).

While, for a given $\alpha_{\rm PC}$, PCMCI+ can have higher adjacency recall, the fair comparison here is the area under the precision-recall curve, which is higher for Bagged-PCMCI+. This implies that one can always choose a higher $\alpha_{\rm PC}$ to obtain a better recall with Bagged-PCMCI+, while still retaining the same or better precision.

\clearpage

\subsection{Nonlinear and mixed noise experiments with Bagged-PCMCI+}\label{nonlinear_mixed}

\begin{figure}[ht]
  \includegraphics[width=0.495\textwidth]{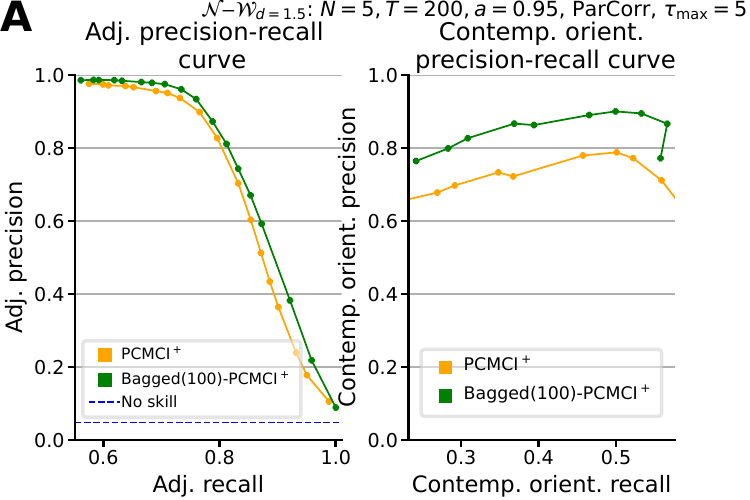}
  \includegraphics[width=0.495\textwidth]{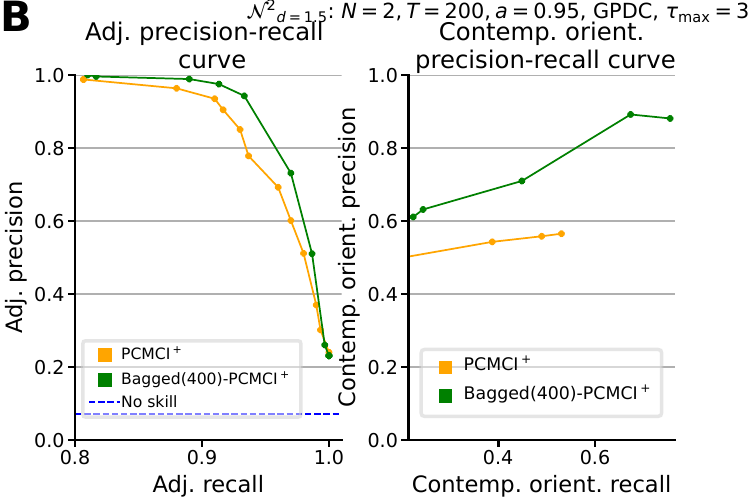}
  \centering
  \caption{(\textbf{A}) Precision-recall curves for adjacencies (left) and contemporaneous orientations (right) obtained by varying the significance level \(\alpha_{PC}\) in PCMCI+ (orange line) and Bagged-PCMCI+ with 100 bootstrap replicas (green line) for the model setup as shown in the top right. In particular, the model is \textbf{linear with mixed noise}: 50\% Gaussian and 50\% Weibull. (\textbf{B}) Precision-recall curves for adjacencies (left) and contemporaneous orientations (right) obtained by varying the significance level \(\alpha_{PC}\) in PCMCI+ (orange line) and Bagged-PCMCI+ with 400 bootstrap replicas $B$ (green line) for a \textbf{nonlinear} model ($f_i(x) = x + 5 x^2 e^{-x^2/20}$, similar to Runge et al., (2020)) with Gaussian noise.}
  \label{fig:nonlinear_mixed}
\end{figure}

In this subsection, we consider two model setups: linear with mixed noise (50\% Gaussian and 50\% Weibull), and nonlinear (50\% linear and 50\% nonlinear dependencies) with Gaussian noise. Results are shown in \textbf{ Figure \ref{fig:nonlinear_mixed}A} and \textbf{ Figure \ref{fig:nonlinear_mixed}B} respectively. Similar to the results of the linear Gaussian setup, the PR curves of Bagged-PCMCI+ systematically dominate the PR curves of the base PCMCI+ algorithm. This confirms the improvement of Bagged-PCMCI+ compared to PCMCI+ in terms of precision and recall for the nonlinear and mixed noise model setups.

\clearpage

\subsection{Additional Bagged-PCMCI+ confidence measure evaluation}

\begin{figure}[h]
  \centering
  \includegraphics[width=0.5\linewidth]{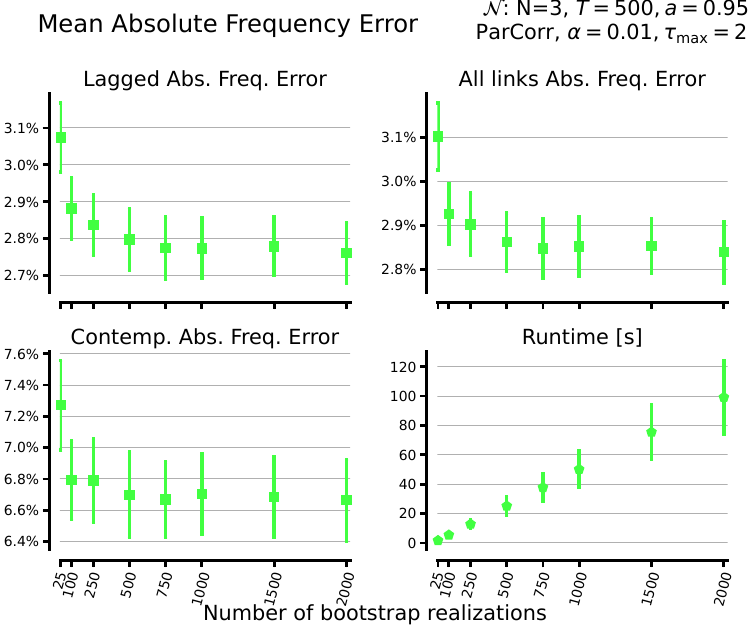}
  \caption{Mean absolute frequency error between estimated reference link frequencies and Bagged-PCMCI+ link frequencies for varying \(B\) and a linear Gaussian setup with parameters indicated at the top right.
  }
  \label{truevbootfreq2}
\end{figure}

\begin{figure}[ht]
  \centering
  \includegraphics[width=0.73\linewidth]{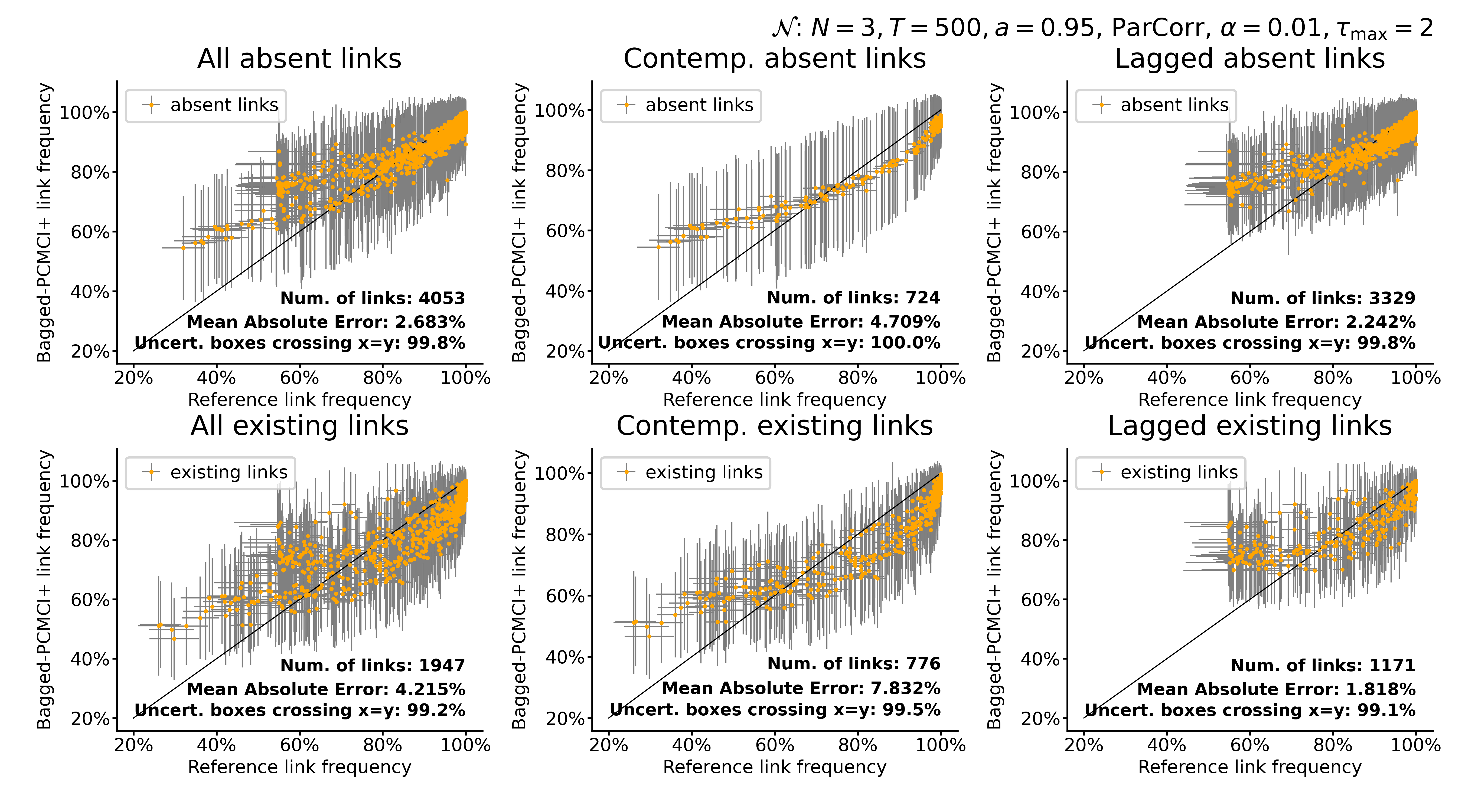}\\
  \includegraphics[width=0.73\linewidth]{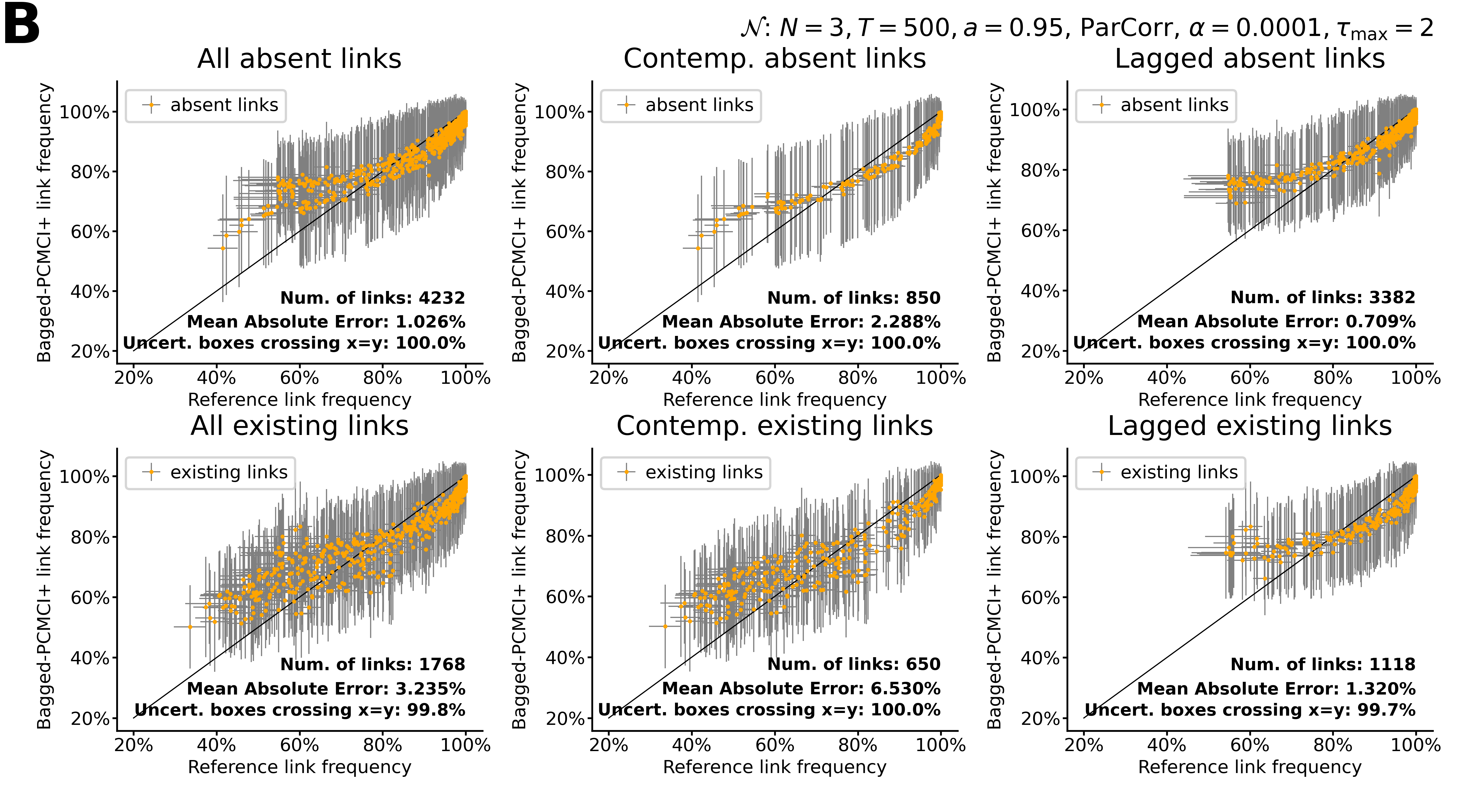}\\
  \includegraphics[width=0.73\linewidth]{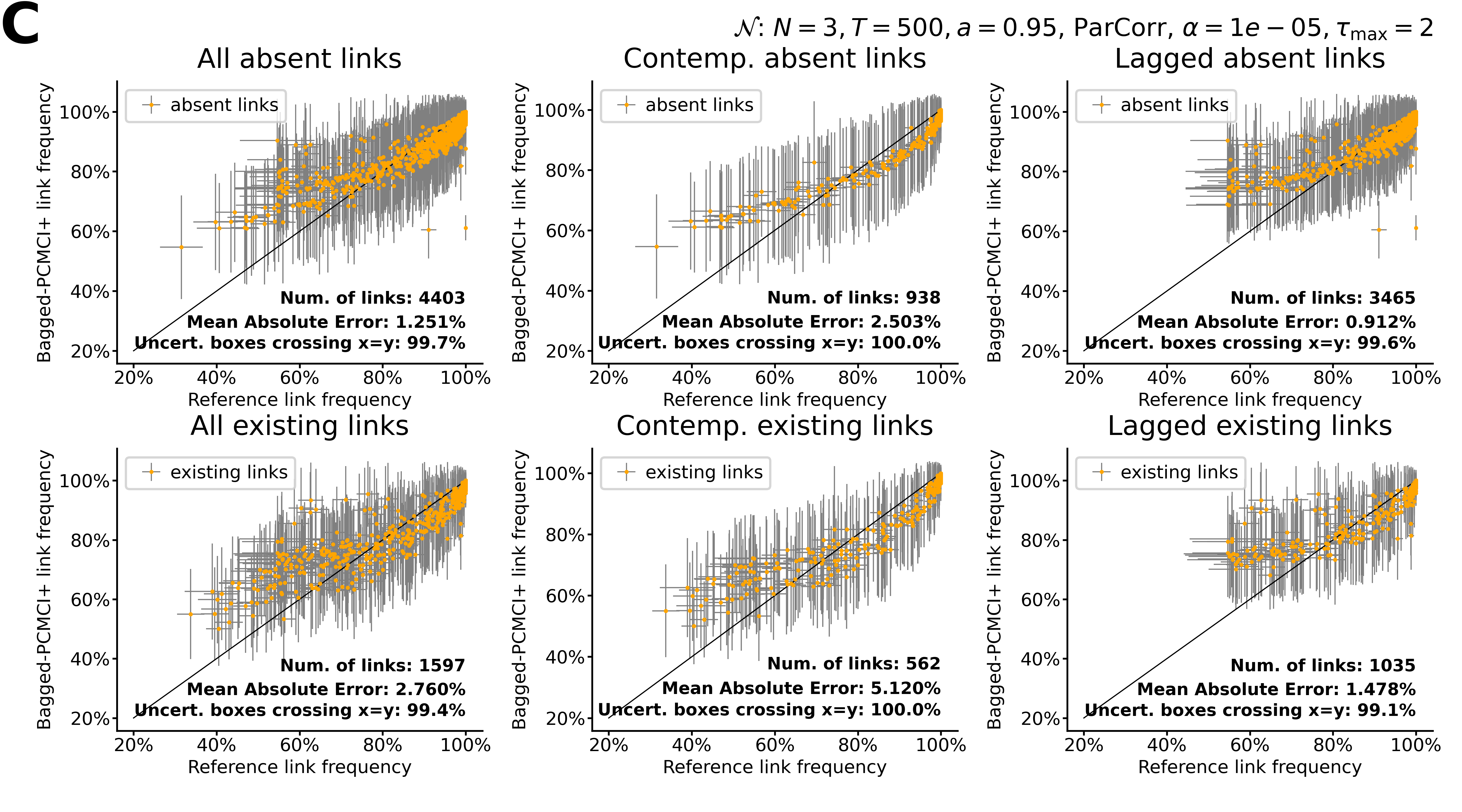}
  \caption{Estimated reference link frequencies against mean Bagged-PCMCI+ link frequencies (\(B=250\)) for a linear Gaussian setup with parameters indicated at the top right. Grey bars indicate the one standard deviation error bars around the estimated value. The same model parameters are used in all three subfigures. Only the significance level $\alpha_{PC}$ changes: (\textbf{A}) $\alpha_{PC}=0.01$, (\textbf{B}) $\alpha_{PC}=10^{-4}$, (\textbf{C}) $\alpha_{PC}=10^{-5}$.}
  \label{sm_true_vs_boot_freq}
\end{figure}

Below, we evaluate our proposed confidence measures for varying significance level $\alpha_{PC}$ to study whether the bootstrapped confidence estimates approximate the estimated reference link frequencies for $\alpha_{PC}\to 0$ (\textbf{Figure \ref{sm_true_vs_boot_freq}}). To reduce computational time, the setup here was slightly modified compared to the main text. While we used $B=1000$ and $L=3$ (number of cross-links) in the main body of the paper, here we set $B=250$ and $L=5$. 
We vary $\alpha_{PC}$ from $0.01$ to $10^{-5}$ to study the mean absolute error between the bootstrapped confidence estimates and the estimated reference link frequencies.

\begin{table}[ht]
\caption{Mean absolute error (in \%) between the bootstrapped confidence estimates and the estimated reference link frequencies (extracted from Figure ~\ref{sm_true_vs_boot_freq}).}
\label{tab:confidence}
\begin{tabular}{l|ccc}
                        & $\alpha_{\rm PC}=10^{-2}$ & $\alpha_{\rm PC}=10^{-4}$ & $\alpha_{\rm PC}=10^{-5}$ \\
                        \hline
All absent links        & 2.7                       & 1.0                       & 1.3                       \\
All existing links      & 4.2                       & 3.2                       & 2.8                       \\
Contemp. absent links   & 4.7                       & 2.3                       & 2.5                       \\
Contemp. existing links & 7.8                       & 6.5                       & 5.1                       \\
Lagged absent links     & 2.2                       & 0.7                       & 0.9                       \\
Lagged existing links   & 1.8                       & 1.3                       & 1.5                      
\end{tabular}
\end{table}

We summarize the results regarding mean absolute error in Tab.~\ref{tab:confidence}. There does seem to be a decrease in error from $\alpha_{\rm PC}=10^{-2}$ to $\alpha_{\rm PC}=10^{-4}$ across all types of link frequencies, while there are mixed results for $\alpha_{\rm PC}=10^{-5}$. There is a visible recurrent positive bias for low values of the true frequencies (approximately 40-60\%): The bootstrapped confidence measures tend to consistently overestimate the reference link frequencies for this range. 
More research is needed to clarify whether the bootstrap confidence estimates do approximate the reference link frequencies, or whether there are persistent biases. In this case, a question would be what this bias depends on (number of variables, graph structure, SCM properties, sample size, etc).

\clearpage

\section{Experiments for Bagged-PC}\label{section:pc}

The numerical results with Bagged-PCMCI+ have shown that the bagging approach leads to enhanced precision-recall when paired with PCMCI+. To demonstrate that this conclusion not only applies to PCMCI+ but also to other causal discovery methods, we carried out further experiments, here with the PC algorithm. We combined our bagging approach with the PC algorithm (referred to as Bagged-PC) and compared its performance against the base PC algorithm. Both the base PC algorithm and Bagged-PC are adapted to time series as given in \citet{Runge2020DiscoveringCA}.

\begin{figure}[ht]
  \includegraphics[width=0.7\linewidth]{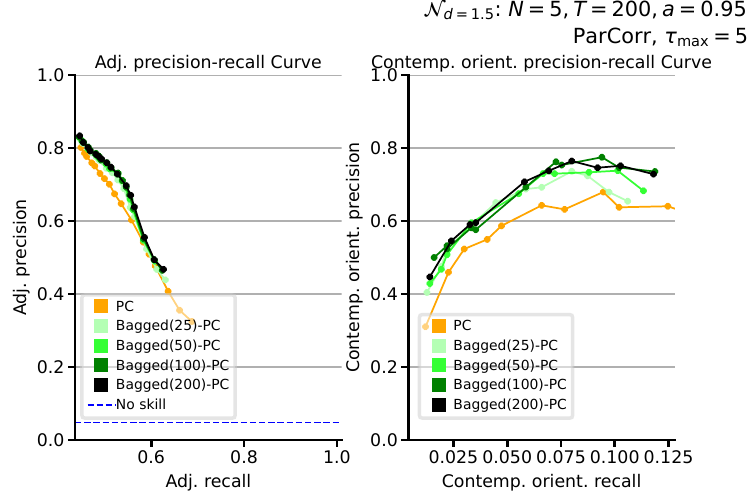}
  \centering
  \caption{Precision-recall curves for adjacencies (left) and contemporaneous orientations (right) obtained by varying the significance level \(\alpha_{PC}\) in \textbf{PC} and \textbf{Bagged-PC} for the model setup as shown in the header. Results are shown for PC (orange line) and Bagged-PC with different numbers of bootstrap replicas $B$ (lines with different shades of green). Here the PC algorithm is adapted for time series data.}
  \label{pcalg_prcurve}
\end{figure}

\begin{figure}[ht]
  \includegraphics[width=0.605\linewidth]{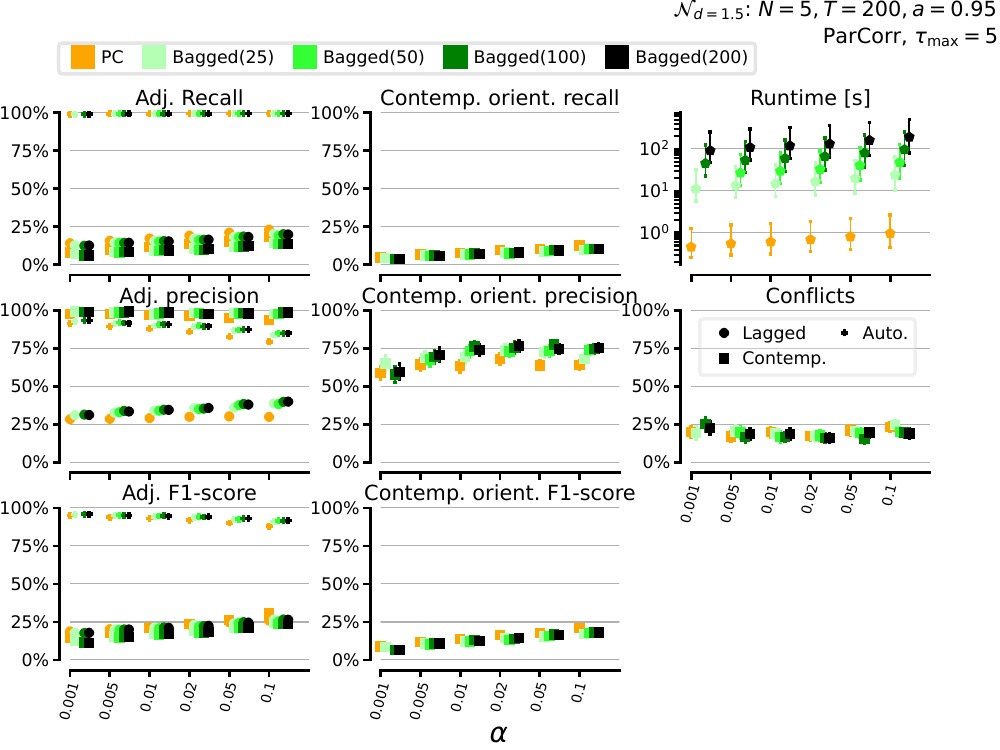}
  \includegraphics[width=0.33\linewidth]{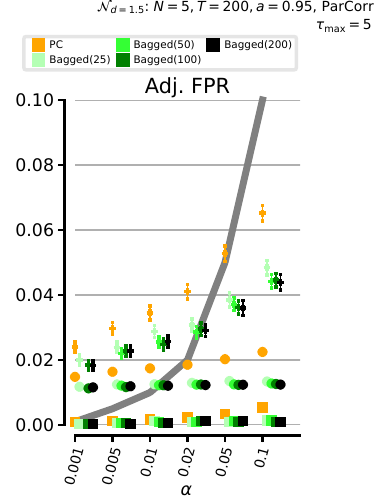}
  \centering
  \caption{Numerical experiments with linear Gaussian setup for a varying \(\alpha_{PC}\) of PC. Here \(N=5\), \(T=200\), and \(a =0.95\).}
  \label{pcalg_panel}
\end{figure}

The results demonstrate that the gain using a bagging approach is similar here: Bagged-PC shows lower FPR and higher precision-recall compared to PC, especially for contemporaneous orientations (see  \textbf{Figure \ref{pcalg_prcurve}} and \textbf{Figure \ref{pcalg_panel}}). 
Hence, our results show that combining a causal discovery method with our bagging approach can considerably improve the performance compared to the base causal discovery method, albeit at the expense of increased computational runtime (if not parallelized).

\clearpage

\section{Experiments for Bagged-LPCMCI}\label{section:lpcmci}

We present here the experiment comparing Bagged-LPCMCI with the base LPCMCI algorithm \citep{NEURIPS2020_94e70705}. LPCMCI is a further development of PCMCI+ that allows for unobserved (hidden/latent) variables. In terms of output, LPCMCI yields a partial ancestral graph (PAG, see \citet{aliMarkovEquivalenceAncestral2009} and \citet{Zhang2008OnTC}) while PCMCI+ outputs a completed partially directed acyclic graph (CPDAG, see \citet{Spirtes2000-SPICPA-2}). A PAG is a causal graph adapted to the presence of latent variables. Relative to a CPDAG, a PAG introduces the additional edge types \(\heado\),\(\ohead\) and \(\headhead\). Here, \(\headhead\) indicates that there is a latent confounder causing both variables. The circle $\circ$ in \(\heado\) and $\ohead$ indicate the uncertainty about the correct causal direction. For example, \(\ohead\) could be $\tailhead$ or $\headhead$ in the true graph. The design of LPCMCI is based on information-theoretical arguments showing that the effect sizes of (conditional) independence tests can often be increased if causal parents of the respective variables are included in the conditioning sets. To utilize this finding to improve recall, the algorithm intertwines the edge-removal phase (finding the graph's skeleton) with the edge-orientation phase and utilizes specific orientation rules to find causal parents and ancestors already before the final skeleton has been found.

\begin{figure}[ht]
  \includegraphics[width=0.56\textwidth]{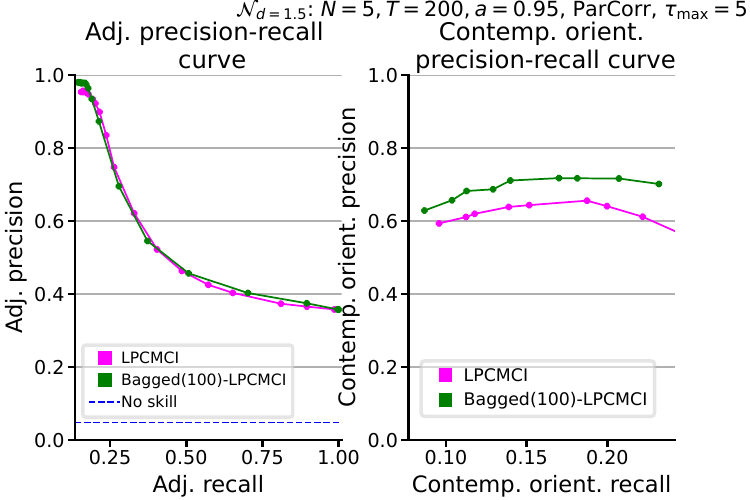}
  \centering
  \caption{Precision-recall curves for adjacencies (left) and contemporaneous orientations (right) obtained by varying the significance level \(\alpha_{PC}\) in \textbf{LPCMCI} (purple line) and \textbf{Bagged-LPCMCI} with 100 bootstrap replicas $B$ (green line) for the model setup as shown in the top right. The hyperparameter $k$ of LPCMCI is set to $4$.}
  \label{fig:LPCMCI}
\end{figure}

The model setup considered here is linear with Gaussian noise, and results are shown in \textbf{Figure \ref{fig:LPCMCI}}. The results demonstrate the gain using a bagging approach: Bagged-LPCMCI shows lower FPR and higher precision-recall regarding contemporaneous orientations compared to the base LPCMCI. For this experiment, we have not found an improvement in the precision-recall of Bagged-LPCMCI regarding adjacencies. Hence, also for LPCMCI, our results show that combining a causal discovery method with our bagging approach can considerably improve the performance compared to the base causal discovery method, albeit at the expense of increased computational runtime (if not parallelized).

\begin{figure}[ht]
  \includegraphics[width=0.605\linewidth]{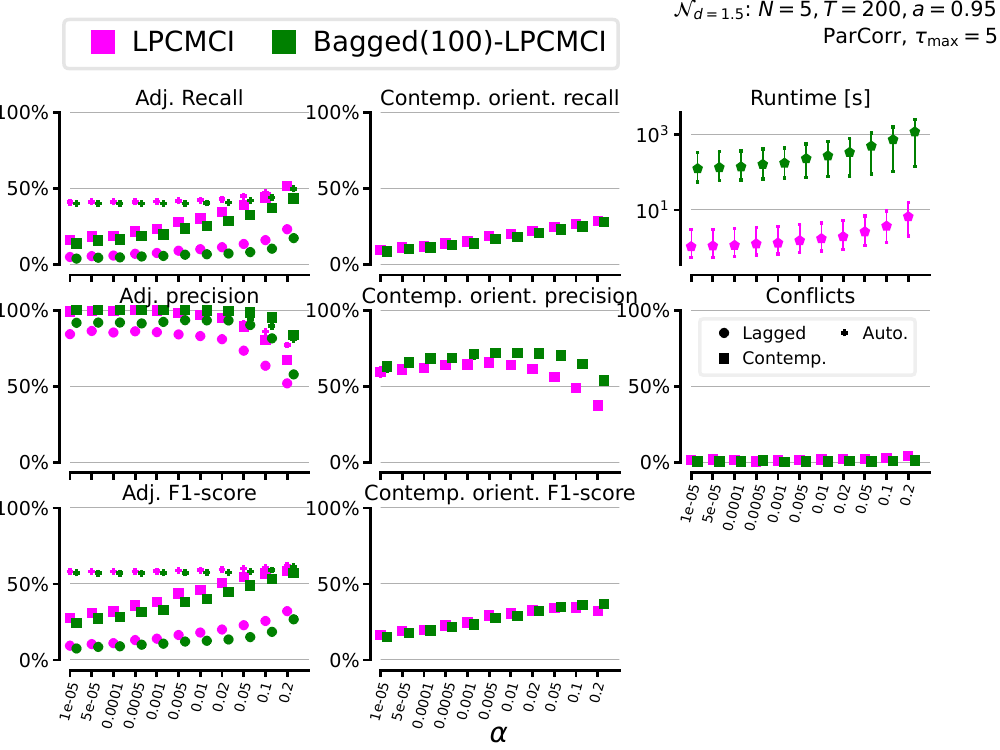}
  \includegraphics[width=0.33\linewidth]{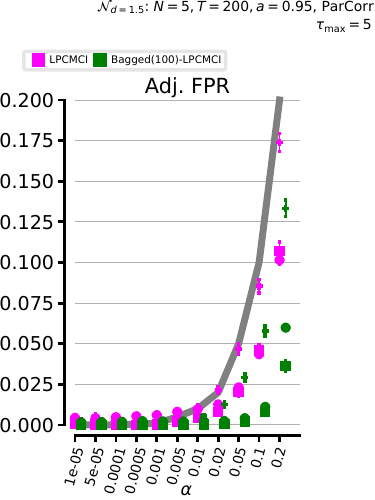}
  \centering
  \caption{Numerical experiments with linear Gaussian setup for a varying \(\alpha_{PC}\) of LPCMCI. Here \(N=5\), \(T=200\), and \(a =0.95\). The hyperparameter $k$ of LPCMCI is set to $4$.}
  \label{lpcmci_panel}
\end{figure}

\end{document}